\documentclass[journal,twoside]{IEEEtran}
\IEEEoverridecommandlockouts

\usepackage{cite}
\usepackage{amsmath,amssymb,amsfonts}
\usepackage{algorithmic}
\usepackage{graphicx}
\usepackage{textcomp}
\usepackage{xcolor,multirow,multicol}
\def\BibTeX{{\rm B\kern-.05em{\sc i\kern-.025em b}\kern-.08em
    T\kern-.1667em\lower.7ex\hbox{E}\kern-.125emX}}
    
\newcommand{\bc}[1]{\mbox{\boldmath $\mathcal{#1}$}}
\newcommand{\mb}[1]{\mathbf{#1}}
\newcommand{\F}{\mathrm{F}}
\newcommand{\T}{\mathrm{T}}

\def\Rb{\mathbb{R}}
\def\A{\mathbf{A}}
\def\B{\mathbf{B}}
\def\C{\mathbf{C}}
\def\a{\mathbf{a}}
\def\b{\mathbf{b}}
\def\c{\mathbf{c}}
\def\t{\mathbf{t}}

\def\Y{\boldsymbol{\mathcal{Y}}}

\def\y{\boldsymbol{y}}
\def\N{\mathcal{N}}
\def\0{\mathbf{0}}

\def\Z{\mathbf{Z}}
\def\zet{\boldsymbol{\zeta}}
\def\delt{\boldsymbol{\delta}}
\def\ro{\boldsymbol{\rho}}
\def\IG{\mathcal{IG}}
\def\G{\mathcal{G}}
\def\I{\mathbf{I}}
\def\GIG{\mathcal{GIG}}
\def\Rho{\boldsymbol{\rho}}
\def\delt{\boldsymbol{\delta}}

\usepackage{tikztensor}
\definecolor{kulblue}{RGB}{0,85,165}
\tikzset{fancy/.style={inner color=kulblue!5!white, outer color=kulblue!20!white, fill faces, opacity=0.95}}
\colorlet{kulblue20}{kulblue!20!white}
\colorlet{kulblue70}{kulblue!70!white}
\colorlet{kulblue30}{kulblue!30!black}
\colorlet{kulblue60}{kulblue!60!black}
\colorlet{kulblue90}{kulblue!90!black}

\begin{document}

\title{Block-Term Tensor Decomposition Model Selection and Computation: The Bayesian Way}

\author{Paris~V.~Giampouras\thanks{P. V. Giampouras is supported by the European Union Horizon~2020 Marie Sk{\l}odowska-Curie Global Fellowship program: HyPPOCRATES—H2020-MSCA-IF-2018, Grant Agreement Number: 844290.}, Athanasios~A.~Rontogiannis,~\IEEEmembership{Senior Member,~IEEE,}
        and Eleftherios~Kofidis,~\IEEEmembership{Member,~IEEE}%
\thanks{P. V. Giampouras is with the Mathematical Institute for Data Science, Johns Hopkins University, Baltimore, MD 212~18, USA. E-mail: 
parisg@jhu.ed}%
\thanks{A. A. Rontogiannis is with the IAASARS, National Observatory of Athens, 152~36 Penteli, Greece. E-mail: tronto@noa.gr.}%
\thanks{E. Kofidis is with the Dept. of Statistics and Insurance Science, University of Piraeus, 185~34 Piraeus, Greece and the Computer Technology Institute \& Press ``Diophantus" (CTI), 265~04 Patras, Greece. E-mail: kofidis@unipi.gr}% 
}

\maketitle

\begin{abstract}
The so-called block-term decomposition (BTD) tensor model, especially in its rank-$(L_r,L_r,1)$ version, has been recently receiving increasing attention due to its enhanced ability of representing systems and signals that are composed of \emph{blocks} of rank higher than one, a scenario encountered in numerous and diverse applications. Uniqueness conditions and fitting methods have thus been thoroughly studied. Nevertheless, the challenging problem of estimating the BTD model structure, namely the number of block terms, $R$, and their individual ranks, $L_r$, has only recently started to attract significant attention, mainly through regularization-based approaches which entail the need to tune the regularization parameter(s). In this work, we build on ideas of sparse Bayesian learning (SBL) and put forward a fully automated Bayesian approach. Through a suitably crafted multi-level \emph{hierarchical} probabilistic model, which gives rise to heavy-tailed prior distributions for the BTD factors, structured sparsity is \emph{jointly} imposed. Ranks are then estimated from the numbers of blocks ($R$) and columns ($L_r$) of non-negligible energy. Approximate posterior inference is implemented, within the variational inference framework. The resulting iterative algorithm completely avoids hyperparameter tuning, which is a significant defect of regularization-based methods. Alternative probabilistic models are also explored and the connections with their regularization-based counterparts are brought to light with the aid of the associated maximum a-posteriori (MAP) estimators. We report simulation results with both synthetic and real-word data, which demonstrate the merits of the proposed method in terms of both rank estimation and model fitting as compared to state-of-the-art relevant methods.% in comparison with alternative Bayesian inference methods and its deterministic, regularization-based counterpart.
\end{abstract}


\section{Introduction}
\label{sec:intro}

\IEEEPARstart{B}{lock-Term Decomposition (BTD)} was introduced in~\cite{ldl08b} as a tensor model that combines the Canonical Polyadic Decomposition (CPD) and the Tucker decomposition (TD)~\cite{sdfhpf17}, in the sense that it decomposes a tensor in a sum of tensors (block terms) that have low multilinear rank
(not necessarily of rank one as in CPD). Hence a BTD can be seen as a constrained TD, with its core tensor being block diagonal (see~\cite[Fig.~2.3]{ldl08b}). It can also be seen as a constrained CPD having factors with (some) collinear columns~\cite{ldl08b}. In a way, BTD lies between the two extremes (in terms of core tensor structure), CPD and TD, and it is useful to recall here the related remark made in~\cite{ldl08b}, namely that ``the rank of a higher-order tensor is actually a combination of the two aspects: one should specify the number of blocks \emph{and} their size". 
Accurately and efficiently estimating these numbers for a given tensor, via a probabilistic approach that relaxes the requirement for hyperparameters tuning, is the main subject of this paper.

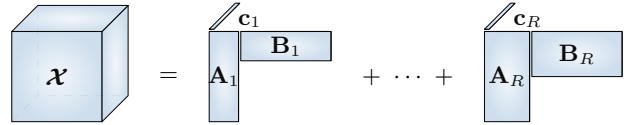
\begin{figure}
\centering
\begin{tikzpicture}[node distance=0.3cm, chain,
term/.style={dim={4,4,3},fancy,tensor scale=0.3}]
\small
% Tensor
\node [tensor, term, dashed back lines] {$\bc{X}$};
% Equality
\node {$=$};
% Term 1
\node [rank-LL1 tensor, term, L=1.3, 2d,
labels={$\mathbf{A}_{1}$}{$\mathbf{B}_{1}$}{$\mathbf{c}_{1}$}] {};
% Plus
\node {$+\ \cdots\ +$};
% Term 2
\node [rank-LL1 tensor, term, L=2, 2d,
labels={$\mathbf{A}_{R}$}{$\mathbf{B}_{R}$}{$\mathbf{c}_{R}$}] {};
\end{tikzpicture}
\caption{Rank-$(L_r,L_r,1)$ block-term decomposition.}
\label{fig:BTD}
\end{figure}

Although~\cite{ldl08b} introduced BTD as a sum of $R$ rank-$(L_r,M_r,N_r)$ terms ($r=1,2,\ldots,R$) in general, the special case of rank-$(L_r,L_r,1)$ BTD has attracted a lot more of attention, because of both its more frequent occurrence in a wide range of applications and the existence of more concrete and easier to check uniqueness conditions (cf.~\cite{rkg21} for an extensive review). This special yet very popular BTD model is at the focus of the present work. Consider a 3rd-order tensor, $\bc{X}\in\mathbb{C}^{I\times J\times K}$. Then its rank-$(L_r,L_r,1)$ decomposition is written as
\begin{equation}
\bc{X}=\sum_{r=1}^{R}\mb{E}_{r}\circ \mb{c}_{r},
\label{eq:BTD1}
\end{equation}
where $\mb{E}_{r}$ is an $I\times J$ matrix of rank $L_r$, $\mb{c}_{r}$ is a nonzero column $K$-vector and $\circ$ denotes outer product. 
Clearly, $\mb{E}_{r}$ can be written as a matrix product $\mb{A}_{r}\mb{B}_{r}^{\T}$ with the matrices $\mb{A}_{r}\in\mathbb{C}^{I\times L_r}$ and $\mb{B}_{r}\in\mathbb{C}^{J\times L_r}$ being of full column rank, $L_r$. Eq.~(\ref{eq:BTD1}) can thus be re-written as 
\begin{equation}
\bc{X}=\sum_{r=1}^{R}\mb{A}_{r}\mb{B}_{r}^{\T}\circ \mb{c}_{r}.
\label{eq:BTD}
\end{equation}
A schematic representation of the rank-$(L_r,L_r,1)$ BTD is given in Fig.~\ref{fig:BTD}.
The $r$th term of this decomposition is a tensor whose frontal slices are all scalar multiples (with the entries of $\mb{c}_r$) of the low-rank matrix $\mb{A}_r\mb{B}_r^{\T}$. 
It should be apparent from~(\ref{eq:BTD}) and Fig.~\ref{fig:BTD} that CPD results as a special case with all $L_r, r=1,2,\ldots,R$ equal to~1.

In general, $R$ and $L_r$, $r=1,2,\ldots,R$ are assumed \emph{a-priori} known (and it is commonly assumed that all $L_r$ are all equal to $L$, for simplicity). However, unless external information is given (such as in a telecommunications ~\cite{ldl12} or a hyperspectral image unmixing application with given or estimated ground truth~\cite{qxzzt17}), there is no way to know these values beforehand. Although overestimation of the ranks $L_r$s of the block terms has been observed not to be harmful in some blind source separation applications (e.g., \cite{ldl12}), this is not the case in general~\cite{rkg21}. Besides, in addition to increasing the computational complexity, setting $L_r$ too high may hinder interpretation of the results through letting noise/artifact sources interfere with the desired sources. This holds for $R$ as well, whose choice is known to be more crucial to the obtained performance as it represents the number of ``factors" that generate the data and its over/under-estimation will lead to over/under-fitting, with undesired consequences for the interpretability of the results (cf.~\cite{rkg21} for related references). 

\subsection{Prior art}

It is known that computing the number of rank-1 terms in a CPD model (i.e. the tensor rank) is NP-hard~\cite{hl13}.
Model selection for BTD is clearly even more challenging than in CPD and TD models and has only recently started to be studied (cf.~\cite{rkg21} for an extensive review of heuristic approaches and techniques).  
The most recent contribution of this kind can be found in our work~\cite{rkg21}, where  the latent factors of BTD are recovered by solving a regularized minimization problem, namely,
\begin{align}
&\underset{\mb{A},\mb{B},\mb{C}}{\mathrm{min}}\frac{1}{2}\left\|\bc{Y} - \sum_{r=1}^R\mb{A}_r\mb{B}_r^{\T}\circ\mb{c}_r\right\|_{\F}^2 + \nonumber \\
 &\lambda\sum_{r=1}^{R}\sqrt{\sum_{l=1}^{L}\sqrt{\|\mb{a}_{r,l}\|_{2}^2 + \|\mb{b}_{r,l}\|_{2}^2 + \eta^2} + \|\mb{c}_r\|_{2}^2 + \eta^2},
\label{eq:minp} 
\end{align}
where $R,L$ are over-estimates of the rank and block ranks of the sought BTD model, $\mb{a}_{r,l},\mb{b}_{r,l}$, $l=1,2,\ldots,L$ are the $l$th columns of $\mb{A}_r,\mb{B}_r$, respectively and $\eta^2$ is a small constant used to ensure smoothness at zero.
Note that~\eqref{eq:minp} is composed of the squared Frobenius norm of the error between the data and its BTD representation and an appropriately chosen regularization term whose minimization promotes structured sparsity over the latent factors of the model. The rank, $R$, and the block ranks, $L_r$, are then taken as the number of $\mb{c}_r$s of non-negligible magnitude and the numbers of non-negligible columns of the corresponding blocks, respectively.  Structured sparsity is favored by the regularizer in a hierarchical, two-level manner, which is  tailored to the form of the BTD model. Indeed, the inner sum of square roots is (excluding the smoothing constant) the sum of the $\ell_2$ norms of the columns of $\left[\begin{array}{cc} \mb{A}^{\T}_r &  \mb{B}_r^{\T}\end{array}\right]^{\T}$, for $r=1,2,\ldots,R$. The well-known column sparsity-promoting effect of this $\ell_{1,2}$ norm leads the superfluous columns of both matrices to be driven jointly to zero, thus providing a ``relaxed" way of penalizing the block ranks of the BTD model. In an analogous manner, the outer sum of the regularizer penalizes  jointly the number of nonzero columns of $\mb{C}$ along with the corresponding blocks $\mb{A}_r\mb{B}^{\top}_r$, which coincides with the number of block terms in the model. The hierarchical alternating iterative reweighted least squares algorithm, called BTD-HIRLS, proposed in~\cite{rkg21} to solve the above problem has demonstrated its competence in revealing the true ranks and accurately computing the model parameters, while enjoying computational efficiency and fast convergence. 

Nevertheless,  being a regularization-based method, BTD-HIRLS faces the same challenge that all such methods have to address, namely to appropriately tune the regularization parameter so as to achieve the best possible performance. Although a rough guideline for the parameter selection has been given and utilized in~\cite{rkg21} as a reference point for the trial-and-error search, this is still only a rule of thumb, not completely relieving the algorithm from the need to spend resources on searching for the most appropriate regularization parameter value.

\subsection{The Bayesian way}

One would thus prefer to be able to automatically (not manually) select the value of the regularization parameter or, more generally, discover the columns of the factor matrices that should be kept, in an automatic, completely data-driven manner. Such a possibility is provided by what is known as \emph{sparse Bayesian leaning (SBL)}~\cite{tp2001,pg08} following the \emph{automatic relevance determination (ARD)} approach, first conceived for and applied in sparsifying the weights of a neural network~\cite{n96}. Through this Bayesian perspective, the unknown parameters of the problem are viewed as random quantities and are each associated with a hyperparameter. Prior distributions suitably assigned to each hyperparameter are conducive  to automatically determining the relevance of the associated parameters at inference time.

In the so-called ARD prior, the parameters are independent and zero-mean Gaussian if conditioned on the values of their hyperparameters, which are represented by the corresponding standard deviations. Hence if the hyperparameter is large enough, the parameter is important whereas for a small enough hyperparameter the corresponding parameter should be suppressed, thus revealing the true complexity (rank) of the model. As stated in~\cite{n96}, ``the posterior distributions of these hyperparameters will reflect which of these situations is more probable, in light of the training data." If a parameter is relevant, this will influence the associated  hyperparameter distribution which in turn will make the parameter more important, in an alternating update cycle between the parameter and hyperparameter posteriors.

ARD was first applied in automatic tensor rank learning on multi-way data modeled via TD in~\cite{mh09}. The hyperparameters (inverse powers of factor columns, also known as precisions) were modeled with Gamma priors, giving rise to the so-called Gauss-Gamma (GG) probabilistic model, where the marginal posterior of the parameters turns out to be a Laplacian, with its well-known sparsity-enforcing effect~\cite{pg08}. 
An analogous GG model was adopted in~\cite{zzc15b} for addressing the corresponding problem for incomplete tensors obeying a CPD model. A fully Bayesian inference approach was taken, in contrast to the \emph{maximum a-posteriori (MAP)} estimation approach followed in~\cite{mh09}. The method proposed in~\cite{zzc15b} performs approximate variational inference (VI)~\cite{tlg08,bca17}, in the sense that the posterior densities are found as the closest (in the sense of minimum Kullback-Leibler (KL) divergence) to the true ones that meet the mean-field assumption of statistical independence of all parameters and hyperparameters. VI is known to be generally faster converging than sampling techniques and better suited to large datasets~\cite{bca17}.

The method of~\cite{zzc15b} was later robustified to cope with incomplete tensors with outliers~\cite{zzzca16}. An online version, for tensors that may grow in time in all their modes and in any order, was reported in~\cite{dzlz18}. Since~\cite{zzc15b}, several works on Bayesian tensor model selection and computation have been reported for both CPD (cf.~\cite{ccswt20} and references therein) and other tensor decomposition models including TD, tensor trains (TT), tensor rings (TR), and t-SVD, among others (see, e.g., \cite{zzc15a,s20,xcww20,scp21,zc21}). In \cite{hz19}, a TT decomposition is employed to compress a deep neural network (DNN) during its training.\footnote{In fact, the power of deep learning (in the form of a convolutional neural network trained on (rank,tensor) pairs) was also exploited to learn to estimate the rank of any given tensor in~\cite{zllcz19}, with results that suggest an improvement over Bayesian schemes like~\cite{zzc15b}. Of course, one should also consider, in such a comparison, the well-known lack of interpretability of a trained deep neural network vis-\`{a}-vis the relatively well-understood principles underlying the purely Bayesian approach.} The TT ranks are automatically determined through a Bayesian GG modeling approach which models the powers of the slices of the TT cores by Gamma priors and couples consecutive cores through the product of their associated hyperparameters.

The GG model is generalized in~\cite{zzc15a} for Bayesian TD by replacing the Gamma hyperprior by an inverse Gamma (IG). This results in a multivariate Laplace marginal prior for the parameters, which also leads to a generalized inverse Gaussian (GIG) for the posterior of the sparsity-inducing precision hyperparameters. Similarly with~\cite{hz19}, the core tensor is indirectly coupled with the matrix factors by using the product of these hyperparameters and the noise precision in the core's normal prior. A more recent generalization of the GG model, this time for CPD rank learning, is developed in~\cite{ccswt20} through a Gauss-GIG mixture that leads to a generalized hyperbolic (GH) marginal prior for the CPD factors. GH is known to be very flexible, including several other sparsity-enforcing distributions as special cases~\cite[Table~I]{bnd14}. The value of this generalization is demonstrated by the fact that the resulting VI method outperforms~\cite{zzc15b} for high-rank tensors and/or low signal-to-noise ratio (SNR). It should be noted, however, that the algorithm in~\cite{ccswt20} is developed on the basis of a simplification of the GH distribution (cf.~Section~IV-C), which effectively leads again to a (generalized) Laplacian marginal prior.

\subsection{Our contribution}

In this paper, we also take a Bayesian approach, viewing the unknowns as random variables and tackling the problem as one of Bayesian modeling and inference~\cite{t20}. The idea is again (as in BTD-HIRLS) 
to impose column sparsity jointly on the factors in a hierarchical, two-level manner. This is achieved through a Bayesian hierarchy of priors with sparsity inducing effect, that realize the coupling of the columns of $\mb{C}$ and the $\mb{A}_r,\mb{B}_r$ blocks at the outer level and that between the columns of corresponding blocks at the inner level. Our choices of priors fall in the class of the so-called exponential power distributions with GIG densities (EP-GIG)~\cite{zwlj12}, which include the GG of~\cite{zzc15b} and the Gauss-GIG and GH of~\cite{zzc15a,ccswt20} as special cases. Inspired by earlier work of ours~\cite{grtk17} and in a manner analogous with the way coupling is achieved in~\cite{zzc15a} for TD, we realize the two-level coupling in the BTD model via appropriately defined products of the associated hyperparameters and the noise precision in the conditional priors of the factors. It is shown that, with our choices of priors, conjugacy is maintained, which allows the development of a tractable approximate inference, efficiently performed via VI~\cite{tlg08,bca17} and leading to an iterative algorithm that comprises closed-form updates and is fast converging. Overestimates of $R$ and the $L_r$s are decreased in the course of the algorithm. This is in contrast to the rank incremental or greedy strategies followed in, e.g., \cite{scp21} and~\cite{bppc13}. Thus, $R$ is estimated as the number of columns of $\mb{C}$ of non-negligible energy while the $L_r$'s are found similarly from the columns of the $\mb{A}_r,\mb{B}_r$ blocks. The Bayesian nature of our approach completely avoids the need for parameter tuning. We also present alternative Bayesian models that reflect simplified causal relationships among the latent variables and thus can be used for lending an insight into the incurred regularization effect through the lens of the MAP-based optimization problems. Simulation results with both synthetic and real data are reported, which demonstrate the effectiveness of the proposed scheme in terms of both rank estimation and model fitting and in comparison with BTD-HIRLS. To the best of our knowledge, the present work is the first of its kind for BTD model selection and computation. A preliminary version can be found in~\cite{grk21a}. In a shorter version, this work was accepted for presentation in EUSIPCO-2021~\cite{grk21b}.

\subsection{Organization of the paper}

The rest of this paper is organized as follows. The adopted notation is described in the following subsection. The problem is mathematically stated in Section~\ref{sec:problem}, where useful expressions for the tensor unfoldings are also recalled. A Bayesian model that implements the idea underlying BTD-HIRLS is developed in Section~\ref{sec:Bayes}. The corresponding approximate inference method is presented and analyzed in Section~\ref{sec:VI}. Alternative probabilistic models, that are inspired from deterministic criteria simpler than~\eqref{eq:minp}, are considered in Section~\ref{sec:others} along with the associated MAP estimators, which clarify the connections with the regularization-based approach. Section~\ref{sec:sims} reports and discusses the simulation results. Conclusions are drawn and future work plans are outlined in Section~\ref{sec:concls}. 

\subsection{Notation}
\label{subsec:notation} 

Lower- and upper-case bold letters are used to denote vectors and matrices, respectively. We denote matrix rows with bold italic letters and we use roman letters for the matrix columns. Higher-order tensors are denoted by upper-case bold calligraphic letters. For a tensor $\bc{X}$, $\mb{X}_{(n)}$ stands for its mode-$n$ unfolding. $\ast$ stands for the Hadamard product and $\otimes$ for the Kronecker product. The Khatri-Rao product is denoted by $\odot$ in its general (partition-wise) version and by $\odot_{\mathrm{c}}$ in its column-wise version. $\circ$ denotes the outer product. The superscript $^{\T}$ stands for transposition. The identity matrix of order $N$ and the all ones $M\times N$ matrix are respectively denoted by $\mb{I}_N$ and $\mb{1}_{M\times N}$. $\mb{1}_N$ stands for $\mb{1}_{N\times 1}$. 
$\mathrm{diag}(\mb{x})$ is the diagonal matrix with the vector $\mb{x}$ on its main diagonal. 
The Euclidean vector norm and the Frobenius tensor norm are denoted by $\|\cdot\|_{2}$ and $\|\cdot\|_{\F}$, respectively. $\mathrm{tr}\{\cdot\}$ stands for the trace operator. $\mathcal{N}(\mb{x}|\boldsymbol{\mu},\mb{\Sigma})$ denotes the normal probability density function (pdf) for a random vector $\mb{x}$ with mean $\boldsymbol{\mu}$ and covariance $\mathbf{\Sigma}$. $\mb{x}$ is omitted when it is easily understood from the context. The generalized inverse Gaussian (GIG) pdf~\cite{lc11} is given by $\GIG(x | p,a,b)=\frac{(a/b)^{p/2}\mathrm{exp}\left[(p-1)\log x-\left(ax+\frac{b}{x}\right)/2\right]}{2\mathcal{K}_p(\sqrt{ab})}$, where $x>0$, $p$ is real, and $\mathcal{K}_p(\cdot)$ is the modified Bessel function of the second kind with index $p$. The Gamma pdf with shape $\zeta$ and rate $\tau$ results as a special case for $b\rightarrow 0,p>0$ and is defined as $\G(x | \zeta,\tau)=\frac{\tau^{\zeta}}{\Gamma(\zeta)}x^{\zeta-1}e^{-\tau x}=\exp\left[(\zeta-1)\log x-x\tau-\log\Gamma(\zeta)+\zeta\log\tau\right]$, where $\Gamma(\cdot)$ is the Gamma function, $\Gamma(\zeta)=\int_{0}^{\infty}x^{\zeta-1}e^{-x}dx$. The inverse (or reciprocal) Gamma pdf also results from the GIG one as a special case (for $a \rightarrow 0$, $p<0$) and, in its shape ($\zeta$) and scale ($\tau$) parametrizarion, is given by $\IG(x | \zeta, \tau)=\frac{\tau^{\zeta}}{\Gamma(\zeta)}x^{-(\zeta+1)}e^{-\tau/ x}=\exp\left[-(\zeta+1)\log x-\frac{\tau}{x}-\log\Gamma(\zeta)+\zeta\log\tau\right]$, for $x>0$.
Sets are denoted by calligraphic letters. For a set $\mathcal{M}$, $|\mathcal{M}|$ is its cardinality. $\mathbb{R}$ and $\mathbb{C}$ are the fields of real and complex numbers, respectively. 

\section{Problem Statement}
\label{sec:problem}

Given the $I\times J\times K$ tensor
\begin{equation}
\bc{Y}=\bc{X}+\sigma \bc{N},
\label{eq:Y=X+N}
\end{equation}
where $\bc{X}$ is given by~(\ref{eq:BTD}) and $\bc{N}$ is a $I\times J\times K$ noise tensor of zero-mean unit variance i.i.d. Gaussian entries, with $\sigma$ being the noise standard deviation, we aim at estimating $R$, $L_r, r=1,2,\ldots,R$ and the factor matrices $\mb{A}_r=\left[\begin{array}{cccc} \mb{a}_{r,1} & \mb{a}_{r,2} & \cdots & \mb{a}_{r,L_r} \end{array}\right]\in\mathbb{C}^{I\times L_r}$, 
$\mb{B}_r=\left[\begin{array}{cccc} \mb{b}_{r,1} & \mb{b}_{r,2} & \cdots & \mb{b}_{r,L_r} \end{array}\right]\in\mathbb{C}^{J\times L_r}$, $\mb{C}\in\mathbb{C}^{K\times R}$, subject of course to the inherent ambiguity resulting from the fact that only the product $\mb{A}_r\mb{B}_r^{\T}$ can be uniquely identified modulo a scaling (with a counter-scaling of $\mb{c}_r$)~\cite{ldl08b}. In terms of its mode unfoldings $\mb{X}_{(1)}\in\mathbb{C}^{I\times JK}$, $\mb{X}_{(2)}\in\mathbb{C}^{J\times IK}$ and $\mb{X}_{(3)}\in\mathbb{C}^{K\times IJ}$, the tensor $\bc{X}$ can be written as~\cite{ldl08b} 
\begin{eqnarray}
\mb{X}_{(1)}^{\T} & = & (\mb{B}\odot\mb{C})\mb{A}^{\T}\triangleq \mb{P}\mb{A}^{\T}, \label{eq:X1} \\
\mb{X}_{(2)}^{\T} & = & (\mb{C}\odot\mb{A})\mb{B}^{\T}\triangleq \mb{Q}\mb{B}^{\T}, \label{eq:X2} \\
\mb{X}_{(3)}^{\T} & = & 
\left[\begin{array}{ccc} (\mb{A}_1\odot_{\mathrm{c}} \mb{B}_1)\mb{1}_{L_1} & \cdots & (\mb{A}_R\odot_{\mathrm{c}} \mb{B}_R)\mb{1}_{L_R}\end{array}\right]\mb{C}^{\T} \nonumber\\
&\triangleq& \mb{S}\mb{C}^{\T}.
\label{eq:X3}
\end{eqnarray}

In this paper, we follow a Bayesian approach to address the above problem, starting from overestimates of $R$ and $L_r$, $r=1,2,\ldots,R$.

\section{The Proposed Bayesian Model}
\label{sec:Bayes}

Let $R$ and the $L_r$s be overestimated to $R_{\mathrm{ini}}$ and $L_{\mathrm{ini}}$, respectively. We intend to place heavy-tailed  distributions, known for their sparsity-inducing effect, over the columns of $\mb{A}_r$s, $\mb{B}_r$s, and $\mb{C}$ in a way that implicitly implements a regularization analogous to that of the BTD-HIRLS method~\cite{rkg21}. Namely, the number of block terms and the ranks of $\mb{A}_r$s and $\mb{B}_r$s are jointly penalized, while respecting the different role that these matrices play in the BTD model. This results in the nulling of all but $R$ columns of $\C$, and the nulling of all but $L_r$ columns of the corresponding ``surviving" $\mb{A}_r,\mb{B}_r$ blocks. Following the premise of the ARD framework and building upon ideas of SBL~\cite{tp2001,t20}, the priors are assigned via a 3-level hierarchy of conjugate prior distributions outlined next.

The likelihood function, which encodes the underlying causal relation between the data and the latent variables, can be written in three equivalent forms, with respect to (w.r.t.) the three unfoldings of $\bc{Y}$ (cf.~(\ref{eq:X1}), (\ref{eq:X2}), (\ref{eq:X3})), as follows:
\begin{align}
    p(\mb{Y}_{(1)}^{\T}\mid \A,\B,\C,\beta) &=  \prod^I_{i=1}p(\y_{(1)i}\mid \A,\B,\C,\beta)  \nonumber \\ 
    &= \prod^I_{i=1}\N(\y_{(1)i} \mid \mb{P}\boldsymbol{a}_i,\beta^{-1}\I_{JK}) 
    \label{eq:likhood1}, \\
    p(\mb{Y}_{(2)}^{\T}\mid \A,\B,\C,\beta) & =  \prod^J_{j=1}p(\y_{(2)i}\mid \A,\B,\C,\beta) \nonumber\\
    &  = \prod^J_{j=1}\N(\y_{(2)j} \mid \mb{Q}\boldsymbol{b}_j,\beta^{-1}\I_{IK}),
    \label{eq:likhood2} \\
    p(\mb{Y}_{(3)}^{\T}\mid \A,\B,\C,\beta) & =  \prod^K_{k=1}p(\y_{(3)k}\mid \A,\B,\C,\beta) \nonumber \\
    & = \prod^K_{k=1}\N(\y_{(3)k} \mid \mb{S}\boldsymbol{c}_k,\beta^{-1}\I_{IJ}),
    \label{eq:likhood3}
\end{align}
where $\beta$ is the noise precision and  $\boldsymbol{a}_i,\boldsymbol{b}_j,\boldsymbol{c}_k$ and $\y_{(1)i},\y_{(2)j},\y_{(3)k}$ are the $i$th, $j$th, $k$th rows of $\A,\B,\C$  and $\mb{Y}_{(1)}$, $\mb{Y}_{(2)}$, $\mb{Y}_{(3)}$, respectively, in column form. The matrices $\mb{A},\mb{B},\mb{C}$ are considered as unobserved variables and are assigned 3-level hierarchical prior distributions. At the first level of the hierarchy, Gaussian distributions are placed over $\A$, $\B$, and $\C$, namely, 
\begin{align}
    p(\A\mid \t,\zet,\beta) & =  \prod^I_{i=1}\N(\boldsymbol{a}_i\mid\0,\beta^{-1}\mb{T}^{-1} (\Z^{-1}\otimes \I_{L_{\mathrm{ini}}}), \label{eq:pA} \\
    p(\B\mid\t,\zet,\beta) & =   \prod^J_{j=1}\N(\boldsymbol{b}_j\mid\0,\beta^{-1}\mb{T}^{-1} (\Z^{-1}\otimes \I_{L_{\mathrm{ini}}})), \label{eq:pB} \\
    p(\C\mid \zet,\beta) & =  \prod^K_{k=1}\N(\boldsymbol{c}_k \mid\0,\beta^{-1}\Z^{-1}), \label{eq:pC}
\end{align}
where $\mb{T}=\mathrm{diag}({\t})$ with $\t\in \Rb^{L_{\mathrm{ini}}R_{\mathrm{ini}}\times 1}$ and $\Z=\mathrm{diag}(\zet)$, $\zet\in\Rb^{R_{\mathrm{ini}}\times 1}$. 
Note that the priors of $\A$ and $\B$ are zero-mean with the {\it same covariance matrix}, which is formed from the diagonal matrices $\Z$ and $\mb{T}$. This particular selection is critical from an implicit regularization perspective, since it induces {\it identical sparsity patterns} over columns/sub-blocks of $\A$ and $\B$.  In the next section, we will see how the posterior covariance matrices of the latent BTD factors will determine the redundant block terms and columns of $\mb{A}_{r}, \mb{B}_{r}$s after the inference process. We can thus claim that  $\Z$ and $\mb{T}$ play a similar  role to that of the two components of the regularizer in~\eqref{eq:minp}. Namely, a sufficiently large value of $\zeta_r$ will lead the $r$th column of $\C$ (cf.~(\ref{eq:pC})) and the entire set of the redundant $L_{\mathrm{ini}}$ columns of sub-matrices $\A_r, \B_r$ (cf.~(\ref{eq:pA}), (\ref{eq:pB})) to zero, acting like the outer sum of square roots in~\eqref{eq:minp}. Moreover, the superfluous $l$th columns of the ``surviving" $\A_r, \B_r$ are \emph{jointly} forced to be zero when the value of $t_{r,l}$ becomes sufficiently large (cf.~(\ref{eq:pA}), (\ref{eq:pB})). Hence $\mb{T}$ plays a role similar to that of the inner sum of square roots of the regularizer in~\eqref{eq:minp}.  That being said, the $R_{\mathrm{ini}}$ diagonal values of $\Z$ act as weights that determine the number of nonzero block terms, $R$, while the $L_{\mathrm{ini}}R_{\mathrm{ini}}$ diagonal values of $\mb{T}$ determine the number of nonzero columns of $\A$ and $\B$. Interestingly, $\mb{Z}$ and $\mb{T}$ are learned from data, thus providing a compelling way to perform BTD model selection.

At the second level of the hierarchy of priors, IG priors are assigned over $\mb{t}$ and $\zet$,
\begin{align}
    p(\t) & =  \prod_{r=1}^{R_{\mathrm{ini}}}\prod_{l=1}^{L_{\mathrm{ini}}} \IG\left (t_{r,l}\left | \frac{I+J+1}{2},\frac{\delta_{r,l}}{2}\right.\right), \\
    p(\zet) & =  \prod^{R_{\mathrm{ini}}}_{r=1}\IG\left (\zeta_r \left |\frac{(I + J)L_{\mathrm{ini}}+ KR_{\mathrm{ini}}+1}{2},\frac{\rho_r}{2}\right.\right ),
\end{align}
where $\delta_{r,l}$ and $\rho_r$ are the scale parameters of the distributions over $t_{r,l}$ and $\zeta_r$, respectively. The third level involves Gamma prior distributions over the scale parameters of the IG distributions of $t_{r,l}$ and  $\zeta_r$, namely, 
\begin{align}
p(\delta_{r,l}) = \G(\delta_{r,l}\mid \psi,\tau),\\
      p(\rho_r) = \G(\rho_r \mid \mu,\nu),
\end{align}
where $\psi,\tau,\mu,\nu$ take very small positive values rendering the respective priors non-informative. 

Note that these priors are conjugate w.r.t. the likelihood functions and w.r.t. each other, which guarantees that the posterior distributions will belong to the same class of distributions with the priors~\cite{t20}.  Finally, we assign a Gamma distribution to the noise precision $\beta$ as follows:
\begin{equation}
p(\beta) = \mathcal{G}(\beta \mid \kappa,\theta).
\end{equation}
Similarly to the hyperparameters of the variables $\delta,\rho$, $\kappa$ and $\theta$ are being set to small positive values rendering the prior non-informative, in the sense that the influence of the prior upon conditioning on the data and the inference process becomes negligible.

\begin{figure}
    \centering
    \includegraphics[width=0.5\textwidth]{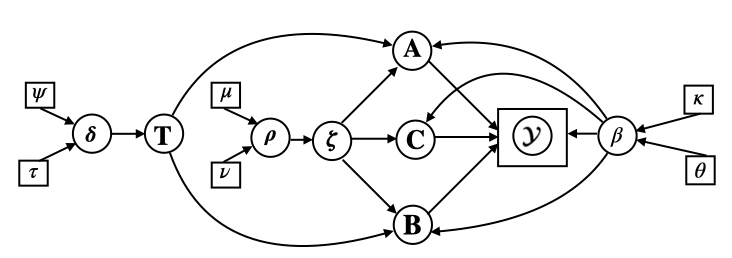}
    \caption{The proposed Bayesian model.}
    \label{fig:graph}
\end{figure}

The adopted Bayesian model is depicted in Fig.~\ref{fig:graph} in the form of a graphical model (with the meaning of $\boldsymbol{\delta},\boldsymbol{\rho}$ being obvious) that manifests the causal relationships of the involved random variables. 
The proposed 3-level hierarchy of priors leads to a heavy-tailed distribution over the columns $\A,\B$ and $\C$, thus allowing for simultaneously learning the latent factors of the BTD model and revealing their ranks. Note that the joint marginal pdf of $\A,\B,\C$ resulting from the hierarchical distributions assigned to the latent factors and their hyperparameters cannot be analytically obtained due to the complexity of the model. Namely, the interrelation of the variables $\A,\B,\C$ with both $\t$ and $\zet$ renders the derivation of their joint pdf an infeasible task. In an effort to provide an insight into the heavy-tailed properties of the distributions assigned over the columns of $\A,\B$ and $\C$, we give in Section~\ref{sec:others} the analytical expression of the joint marginal pdf for a similar but slightly ``relaxed" hierarchical Bayesian model. We would like to stress again at this point that the model described in this section allows us to follow a hyperparameter tuning-free approach since all involved parameters are treated as random variables. The way this is done is detailed next.

\section{Approximate Posterior Inference}
\label{sec:VI}

Let $\boldsymbol{\Theta}$ be the cell array which includes all unobserved variables, that is, $\boldsymbol{\Theta}\triangleq\{ \A,\B,\C,\t,\zet,\beta,\Rho,\delt \}$. The exact joint posterior of the variables of the adopted Bayesian model is given by
\vspace{-0.3cm}
\begin{align}
    p(\boldsymbol{\Theta} \mid \Y) = \frac{p(\Y,\boldsymbol{\Theta})}{\int p(\Y,\boldsymbol{\Theta})d\boldsymbol{\Theta}}.
    \label{eq:post}
\end{align}
Due to the complexity of the model, the marginal distribution of $\Y$ in the denominator is computationally intractable. Therefore, we follow a variational inference (VI) approach for approximating~(\ref{eq:post}). The idea is to approximate the posterior by a distribution which is as close as possible to the exact posterior in terms of the KL divergence~\cite{tlg08}. VI allows for an efficient approximate inference process even in vastly complicated Bayesian models that involve high-dimensional variables. It is usually coupled with mean-field approximation, namely, the assumption that the posterior distribution can be factorized w.r.t. the involved variables, implying statistical independence among them. In our case, the approximate posterior $q(\boldsymbol{\Theta})$ of $p(\boldsymbol{\Theta} \mid \Y)$ is written in the form
\begin{align}
    q(\boldsymbol{\Theta}) &= q(\beta) \prod^I_{i=1}q(\boldsymbol{a}_i)\prod^J_{j=1}q(\boldsymbol{b}_j)\prod^K_{k=1}q(\boldsymbol{c}_k)\times \nonumber \\
    &\prod^{R_{\mathrm{ini}}}_{r=1}\prod^{L_{\mathrm{ini}}}_{l=1}q(t_{r,l})q(\delta_{r,l})\prod^{R_{\mathrm{ini}}}_{r=1}q(\zeta_r)q(\rho_r).
\label{eq:mean-field}
\end{align}
Denoting the individual variables above by $\boldsymbol{\theta}_i$, the corresponding VI-based posteriors are known to satisfy~\cite{tlg08}
\begin{align}
 q(\boldsymbol{\theta}_i) = \frac{\exp(\langle\ln(p(\Y,\boldsymbol{\Theta}))\rangle_{i\neq j})}{\int \exp(\langle\ln(p(\Y,\boldsymbol{\Theta}))\rangle_{i\neq j}d\boldsymbol{\theta}_i}, \label{eq:q}
\end{align}
where $\langle \cdot \rangle_{i\neq j}$ denotes expectation w.r.t. all $q(\boldsymbol{\theta}_j)$s but $q(\boldsymbol{\theta}_i)$.
To solve~(\ref{eq:q}) a block coordinate ascent approach is taken, employing the cyclic update rule, namely solving for $q(\boldsymbol{\theta}_i)$ given $q(\boldsymbol{\theta}_j)$, $j\neq i$ and continuing for all $i$ in a cyclic manner. More specifically, from~(\ref{eq:q}) and using the expression for the likelihood which is based on the mode-1 unfolding of $\bc{Y}$ (cf.~(\ref{eq:likhood1})) the posterior distribution of $\boldsymbol{a}_i$ turns out to be 
\begin{equation}
    q(\boldsymbol{a}_i) = \N(\langle \boldsymbol{a}_i\rangle,\Sigma_{\boldsymbol{a}}), 
    \label{eq:qa}
\end{equation}
with\footnote{All $\boldsymbol{a}_i$'s have the same covariance matrix, $\Sigma_{\boldsymbol{a}}$, and similarly for the $\boldsymbol{b}_j$'s and the $\boldsymbol{c}_k$'s.}
\begin{eqnarray}
 \langle \boldsymbol{a}_i \rangle  & = &  \langle\beta\rangle\Sigma_{\boldsymbol{a}}\langle\mathbf{P}\rangle^{\T}\y_{(1)i}, \\
\Sigma_{\boldsymbol{a}} & = & \langle \beta\rangle^{-1}(\langle \mathbf{P}^{\T} \mathbf{P}\rangle +\langle \mb{T}\rangle(\langle\mb{Z}\rangle\otimes \I_{L_{\mathrm{ini}}}))^{-1},
\end{eqnarray}
where $\langle \cdot \rangle$ denotes expectation w.r.t the posterior of the involved variable.
Now, by employing \eqref{eq:likhood2}, the posterior of $\boldsymbol{b}_j$ results in an analogous manner as:
\begin{equation}
q(\b_j) = \N(\langle \boldsymbol{b}_j\rangle,\Sigma_{\boldsymbol{b}}),
\label{eq:qb}
\end{equation} 
with
\begin{eqnarray}
\langle \boldsymbol{b}_j \rangle & = & \langle\beta\rangle\Sigma_{\boldsymbol{b}}\langle\mathbf{Q}\rangle^{\T}\y_{(2)j} \\
\Sigma_{\boldsymbol{b}} & = & \langle \beta\rangle^{-1}(\langle \mathbf{Q}^{\T} \mathbf{Q}\rangle + \langle \mb{T}\rangle (\langle \mb{Z}\rangle\otimes \I_{L_{\mathrm{ini}}}))^{-1}.
\end{eqnarray}
Concluding the first level of the hierarchy, the posterior of $\boldsymbol{c}_k$ is
\begin{align}
    q(\boldsymbol{c}_k) = \N(\langle \boldsymbol{c}_k\rangle,\Sigma_{\boldsymbol{c}}),
\end{align}
with
\begin{eqnarray}
\langle \boldsymbol{c}_k \rangle & = & \langle\beta\rangle\Sigma_{\boldsymbol{c}}\langle\mathbf{S}\rangle^{\T}\y_{(3)k} \\
\Sigma_{\boldsymbol{c}} & = & \langle\beta\rangle^{-1}(\langle\mathbf{S}^{\T}\mathbf{S}\rangle + \langle\mb{Z}\rangle)^{-1}.
\end{eqnarray}
Next, the approximate posteriors of the variables belonging to the second level of hierarchy are given. Following similar arguments with~\cite{grtk17}, the posterior of $t_{r,l}$ turns out to be a GIG pdf,  
\begin{align}
    q(t_{r,l}) = \GIG\left (t_{r,l}\left | -\frac{1}{2},\langle\beta\rangle\langle\zeta_r\rangle(\langle\a^{\T}_{r,l}\a_{r,l}\rangle + \langle\b^{\T}_{r,l}\b_{r,l}\rangle),\langle\delta_{r,l}\rangle\right.\right),
    \label{eq:qtrl}
\end{align}
with mean 
\begin{align}
    \langle t_{r,l} \rangle = \sqrt{\frac{\langle\delta_{r,l}\rangle}{\langle\beta\rangle\langle\zeta_r\rangle(\langle\a^{\T}_{r,l}\a_{r,l}\rangle+\langle\b^{\T}_{r,l}\b_{r,l}\rangle) }},
\end{align}
where $\langle\a^{\T}_{r,l}\a_{r,l}\rangle$ and $\langle\b^{\T}_{r,l}\b_{r,l}\rangle$ are the $((r-1)L_{\mathrm{ini}}+l,(r-1)L_{\mathrm{ini}}+l)$ entries of 
\begin{equation}
    \langle \A^{\T}\A\rangle=\langle\A\rangle^{\T}\langle\A\rangle + I\Sigma_{\boldsymbol{a}}
    \label{eq:AA}
\end{equation}
and
\begin{equation}
    \langle \B^{\T}\B\rangle= \langle\B\rangle^{\T}\langle\B\rangle + J\Sigma_{\boldsymbol{b}},
    \label{eq:BB}
\end{equation}
respectively.
Similarly, the approximate posterior of $\zeta_r$ is also GIG,
with $\langle \zeta_r \rangle$ given by
\begin{align}
    \langle \zeta_{r} \rangle = \sqrt{\frac{\langle\rho_{r}\rangle}{\langle\beta\rangle(\sum^{L_{\mathrm{ini}}}_{l=1}\langle t_{r,l}\rangle(\langle\a^{\T}_{r,l}\a_{r,l}\rangle+\langle\b^{\T}_{r,l}\b_{r,l}\rangle) + \langle \c^{\T}_r\c_r\rangle)}}
\end{align}
and $\langle \c^{\T}_r\c_r\rangle$ denoting the $(r,r)$ entry of 
\begin{equation}
\langle \C^{\T}\C\rangle=\langle\C\rangle^{\T}\langle\C\rangle + K\Sigma_{\boldsymbol{c}}.
    \label{eq:CC}
\end{equation}
In addition, by employing the pdfs of $t_{r,l}$ and $\zeta_{r}$, the expectations $\langle \frac{1}{t_{r,l}}\rangle$ and $\left\langle \frac{1}{\zeta_r}\right\rangle$, required in the posteriors at the third level of hierarchy, can be expressed as 
\begin{equation}
\left\langle \frac{1}{t_{r,l}} \right\rangle = \frac{1}{\langle \delta_{r,l} \rangle} + \frac{1}{\langle t_{r,l} \rangle},\;\;\;\;  \left\langle \frac{1}{\zeta_{r}} \right\rangle = \frac{1}{\langle \rho_{r} \rangle} + \frac{1}{\langle \zeta_{r} \rangle}.
\end{equation}
Finally, it can be shown (as in~\cite{grtk17}) that, at the third level of hierarchy, the approximate posteriors of $\delta_{r,l},\rho_r$ and $\beta$ are Gamma distributions with $\langle \delta_{r,l}\rangle$, $\langle \rho_{r}\rangle$ and $\langle \beta\rangle$ given in Table~\ref{tab:BBTD}, where the resulting \emph{Bayesian-BTD (BBTD)} algorithm is summarized. 
\begin{table}
\caption{The BBTD algorithm}
\label{tab:BBTD}
\centering
    \begin{tabular}{l}
    \hline
        \textbf{Input}:  $\bc{Y}, R_{\mathrm{ini}}, L_{\mathrm{ini}}$   \\
        \textbf{Output}: $\hat{R}$, $\hat{L}_r, r=1,2,\ldots,\hat{R}$, $\mb{\hat{A}},\mb{\hat{B}},\mb{\hat{C}}$ \\
   \hline
    \textbf{Initialize} $\langle\B\rangle,\langle\C\rangle,\langle\beta\rangle,\langle\mb{T} \rangle,\langle\Z\rangle,\langle\delt\rangle,\langle\Rho\rangle$  \\
    \textbf{repeat} \\
    \;\;\;\;$\Sigma_{\boldsymbol{a}} = \langle \beta\rangle^{-1}(\langle \mathbf{P}^{\T} \mathbf{P}\rangle +\langle \mb{T}\rangle(\langle\mb{Z}\rangle\otimes \I_{L_{\mathrm{ini}}}))^{-1}$  \\
     \;\;\;\;$\langle \A \rangle = \langle\beta\rangle\mathbf{Y}_{(1)} \langle\mathbf{P}\rangle\Sigma_{\boldsymbol{a}}$   \\
     \;\;\;\;$\Sigma_{\boldsymbol{b}} = \langle \beta\rangle^{-1}(\langle \mathbf{Q}^{\T}\mathbf{Q}\rangle + \langle\mathbf{T}\rangle(\langle\mb{Z}\rangle\otimes \I_{L_{\mathrm{ini}}}))^{-1}$\\
     \;\;\;\;$\langle\B\rangle = \langle\beta\rangle\mathbf{Y}_{(2)}\langle\mathbf{Q}\rangle\Sigma_{\boldsymbol{b}}$\\
      \;\;\;\;$\Sigma_{\boldsymbol{c}} = \langle\beta\rangle^{-1}(\langle\mathbf{S}^{\T}\mathbf{S}\rangle + \langle\mb{Z}\rangle)^{-1}$\\
     \;\;\;\;$\langle \C \rangle = \langle\beta\rangle\mathbf{Y}_{(3)}\langle\mathbf{S}\rangle\Sigma_{\boldsymbol{c}}$ \\
     \;\;$r=1,2,\dots,R_{\mathrm{ini}}, l=1,2,\dots,L_{\mathrm{ini}}$ \\
     \;\;\;\; $\langle t_{r,l} \rangle = \sqrt{\frac{\langle\delta_{r,l}\rangle}{\langle\beta\rangle\langle\zeta_r\rangle(\langle\a^{\T}_{r,l}\a_{r,l}\rangle+\langle\b^{\T}_{r,l}\b_{r,l}\rangle) }}$,\;\;\; \\
     \;\;\;\;$\left\langle\frac{1}{t_{r,l}}\right\rangle = \frac{1}{\langle\delta_{r,l}\rangle} + \frac{1}{\langle t_{r,l}\rangle}$,\;\;\;   \\
     \;\;\;\; $\langle\delta_{r,l}\rangle = \frac{2\psi + I+J+1}{2\tau + \langle\frac{1}{t_{r,l}}\rangle}$,\\ 
     \;\;$r=1,2,\dots,R_{\mathrm{ini}}$ \\
     \;\;\;\;$\langle \zeta_{r} \rangle = \sqrt{\frac{\langle\rho_{r}\rangle}{\langle\beta\rangle(\sum^{L_{\mathrm{ini}}}_{l=1}\langle t_{r,l}\rangle(\langle\a^{\T}_{r,l}\a_{r,l}\rangle+\langle\b^{\T}_{r,l}\b_{r,l}\rangle)) + \langle \c^T_r\c_r\rangle }}$\\
      \;\;\;\;$\left\langle\frac{1}{\zeta_{r}}\right\rangle = \frac{1}{\langle\rho_{r}\rangle} + \frac{1}{\langle \zeta_{r}\rangle}$ \\
     \;\;\;\;$\langle\rho_r \rangle = \frac{ 2\mu + (I+J)L_{\mathrm{ini}}+KR_{\mathrm{ini}} + 1}{2\nu + \langle\frac{1}{\zeta_r}\rangle}$ \\
     \;\;\;\;$\langle\beta\rangle = (2\kappa + (I+J)L_{\mathrm{ini}}R_{\mathrm{ini}} + KR_{\mathrm{ini}} + IJK)/ (2\theta + $\\ 
    \;\;\;\; $ \left\langle\left\|\mathbf{Y}_{(1)}^{\T}- \mathbf{P}\A^{\T}\right\|^2_{\F}\right\rangle+\sum^{R_{\mathrm{ini}}}_{r=1}\langle \zeta_r\rangle[\sum^{L_{\mathrm{ini}}}_{l=1}\langle t_{r,l}\rangle(\langle\a^{\T}_{r,l} \a_{r,l}\rangle + $\\
    \;\;\;\; $\langle\b^{\T}_{r,l}\b_{r,l}\rangle) +\langle\c^{\T}_{r} \c_{r}\rangle])$\\
     \textbf{until \textit{convergence}}\\
      $\mathcal{I}=\{i\in\{1,2,\ldots,R_{\mathrm{ini}}\} \mid \mbox{$i$th column of $\langle\C\rangle$ is of non-negligible energy} \}$  \\
     $\hat{R}=|\mathcal{I}|$ \\ 
     $\mb{\hat{C}}=\langle\C\rangle(:,\mathcal{I})$ \\
     Let the elements of $\mathcal{I}$ be sorted in increasing order as $i_1,i_2,\ldots,i_{\hat{R}}$  \\
     $\mathcal{I}_r=\{l\in\{1,2,\ldots,L_{\mathrm{ini}}\} \mid$   \mbox{$l$th column of}  \\
     \mbox{$\langle\A\rangle_r\triangleq\langle\A\rangle(:,(i_r-1)L_{\mathrm{ini}}+1:i_rL_{\mathrm{ini}})$}  $\mbox{is of non-negligible energy}\}$   \\
     $\;\;\;\;\; =\{l\in\{1,2,\ldots,L_{\mathrm{ini}}\} \mid \mbox{$l$th column of}$ \\
     $\mbox{$\langle\B\rangle_r\triangleq\langle\B\rangle(:,(i_r-1)L_{\mathrm{ini}}+1:i_rL_{\mathrm{ini}})$} $ $\mbox{is of non-negligible energy}\}$,  \\
      ($r=1,2,\ldots,\hat{R}$)\\
     $\hat{L}_r=|\mathcal{I}_r|$, $r=1,2,\ldots,\hat{R}$ \\
     $\mb{\hat{A}}_r=\langle\A\rangle_r(:,\mathcal{I}_r)$, 
     $\mb{\hat{B}}_r=\langle\B\rangle_r(:,\mathcal{I}_r)$,  $r=1,2,\ldots,\hat{R}$\\
     $\mb{\hat{A}}=\left[\begin{array}{cccc} \mb{\hat{A}}_1 & \mb{\hat{A}}_2 & \cdots & \mb{\hat{A}}_{\hat{R}} \end{array}\right]$\\
     $\mb{\hat{B}}=\left[\begin{array}{cccc} \mb{\hat{B}}_1 & \mb{\hat{B}}_2 & \cdots & \mb{\hat{B}}_{\hat{R}} \end{array}\right]$\\
     \hline
    \end{tabular}
\end{table}  
The rest of the first- and second-order statistics that are required in the algorithm implementation are computed as in Table~\ref{tab:stats}, based on the assumption of statistically independent $\mb{A},\mb{B},\mb{C}$ (cf.~(\ref{eq:mean-field})) and making use of the identities for the Grammians of Khatri-Rao products proved in~\cite[Appendix~C]{rkg21}.
\begin{table}
\caption{First- and Second-Order Statistics Required in the BBTD Algorithm}
\label{tab:stats}
\centering
\begin{tabular}{|l|}
\hline
$\langle \mb{P} \rangle  =  \langle \B\rangle\odot \langle\C\rangle$\\
 $\langle \mb{Q} \rangle = \langle \C\rangle\odot \langle\A\rangle $\\
  $\langle \mb{S} \rangle   =\left[(\langle \mb{A}_1\rangle \odot_{\mathrm{c}} \langle \mb{B}_1\rangle)\mb{1}_{L_{\mathrm{ini}}}  \cdots  (\langle \mb{A}_{R_{\mathrm{ini}}}\rangle \odot_{\mathrm{c}} \langle \mb{B}_{R_{\mathrm{ini}}}\rangle)\mb{1}_{L_{\mathrm{ini}}}\right]$\\ 
  $\langle \mb{P}^{\T}\mb{P}\rangle = \langle\B^{\T}\B\rangle\ast(\langle\C^{\T}\C\rangle\otimes \mb{1}_{L_{\mathrm{ini}}\times L_{\mathrm{ini}}})$\\
  $\langle \mb{Q}^{\T}\mb{Q}\rangle = \langle\A^{\T}\A\rangle\ast(\langle\C^{\T}\C\rangle\otimes \mb{1}_{L_{\mathrm{ini}}\times L_{\mathrm{ini}}})$ \\
 $\langle \mb{S}^{\T}\mb{S}\rangle  =  (\I_{R_{\mathrm{ini}}} \otimes \mb{1}^{\T}_{L_{\mathrm{ini}}})(\langle \A^{\T}\A\rangle \ast \langle\B^{\T}\B\rangle)(\I_{R_{\mathrm{ini}}} \otimes \mb{1}_{L_{\mathrm{ini}}}) $\\
 $\langle \|\mb{Y}_{(1)}^{\T}-\mb{P}\mb{A}^{\T}\|_{\F}^{2}\rangle=\|\mb{Y}_{(1)}\|_{\F}^{2}-2\mathrm{tr}\{\langle \A\rangle^{\T}\mb{Y}_{(1)}\langle \mb{P}\rangle\}$\\
 $\;\;\;\;\;\;\;\;\;\;+\mathrm{tr}\{\langle \A^{\T}\A \rangle \langle \mb{P}^{\T}\mb{P}\rangle\}$ \\
 \hline
\end{tabular}
\end{table}

$R$ is estimated as the number of columns of $\langle \C \rangle$ of non-negligible energy and similarly for the $L_r$s and the corresponding blocks of $\langle \A \rangle,\langle \B \rangle$, as detailed in Table \ref{tab:BBTD}. 
The iterations stop when a convergence criterion is met (e.g., the relative difference of the tensor reconstruction errors in two consecutive iterations becomes less than a user-defined threshold) or the maximum number of iterations is reached.\noindent

The algorithm can be randomly initialized and, as empirically demonstrated in Section~\ref{sec:sims}, it converges fast and is very robust to initialization. Moreover, in view of its mean-field VI nature, the method is guaranteed to converge to a stationary point of the KL divergence function.

As far as the computational complexity of the algorithm is concerned, the computational cost of a BBTD iteration is similar to that of BTD-HIRLS (cf.~\cite[Appendix C]{rkg21}), with $\mathcal{O}((I+J)L^2+K)R^2)$ extra multiplications required to compute $\langle t_{r,l}\rangle$, $\langle \delta_{r,l}\rangle$, $\langle \zeta_r\rangle$ and $\langle \rho_r\rangle$. $\mathcal{O}(IJK+IJKLR+I(LR)^2+(LR)^3+LR+R)$ additional multiplications are needed in the computation of $\langle \beta \rangle$. Therefore, as in BTD-HIRLS, and for the more realistic case of tensors with dimensions much larger than $R$ and $L$, the number of multiplications required per iteration of BBTD is $\mathcal{O}(IJKLR)$, i.e., of the same order with the computational cost of a BTD-HIRLS iteration~\cite{rkg21}. Clearly, $R$ and $L$
 here refer to their overestimates, $R_{\mathrm{ini}}$ and $L_{\mathrm{ini}}$, respectively. The cost  can be reduced if \emph{pruning} of the nulled columns of $\langle \C\rangle$ and the corresponding blocks of $\langle \A\rangle$ and $\langle \B \rangle$ in the course of the algorithm is included. For the sake of the simplicity of presentation, this is only performed in Table~\ref{tab:BBTD} at the end of the iterative inference.
 
\section{Alternative Bayesian Models and Their Regularization-based Counterparts}
\label{sec:others}

\begin{figure}
\begin{tabular}{c}
    \includegraphics[height=0.2\textwidth,width=0.3\textwidth]{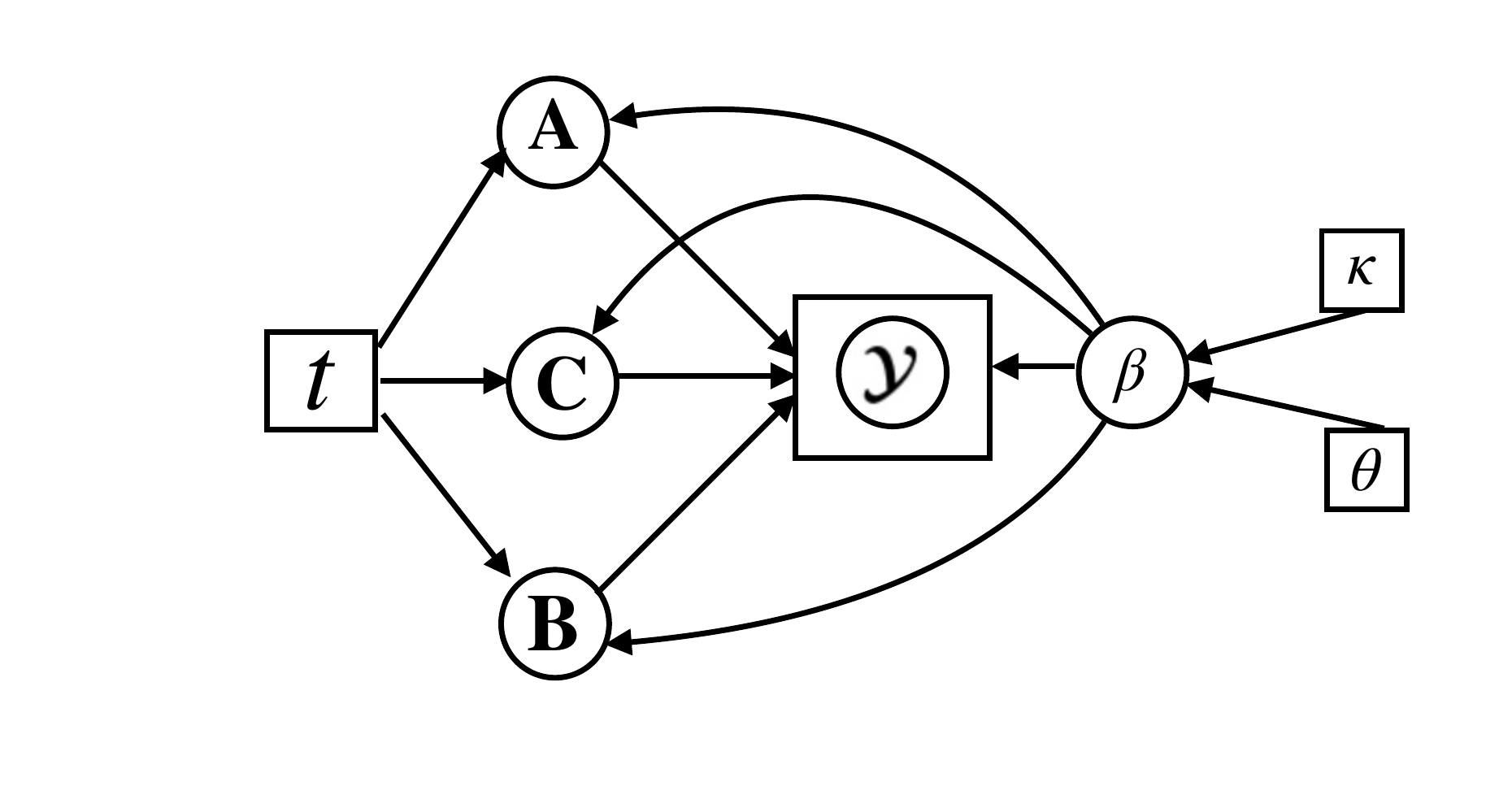} \\
    I \\ \hline
    \includegraphics[height=0.2\textwidth,width=0.45\textwidth]{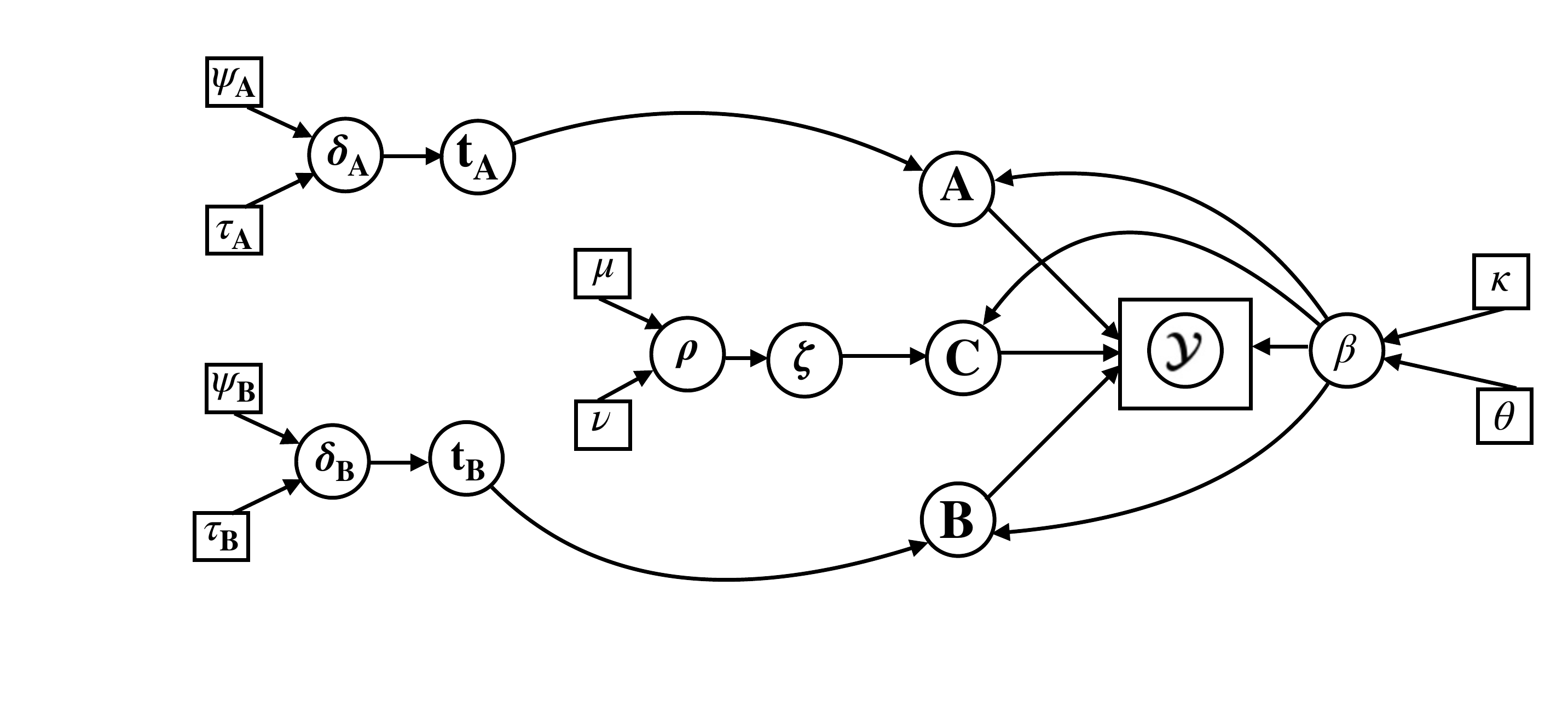}\\
     II
\end{tabular}
    \caption{Alternative probabilistic models.}
    \label{fig:graphs}
\end{figure}

In this section, we present two alternative Bayesian models, which can be viewed as simplified versions of the model introduced in Section \ref{sec:Bayes}. The main goal here is to manifest the role that specific aspects of the adopted model (e.g., the number of the levels in the hierarchy) play in the regularization that is induced to the latent BTD factors at inference time. Both models presented next assume the same likelihood function with the more composite model presented previously. That being said, the main difference between the two models lies in the priors placed over $\A,\B,\C$, as detailed next.
\paragraph{Model~I}
This model consists of a single level of hierarchy, with Gaussian priors being assigned to the rows of $\mb{A},\mb{B}$ and $\mb{C}$:
\begin{eqnarray}
    p(\A | t,\beta) & = & \prod^I_{i=1}\N(\boldsymbol{a}_i\mid\0,\beta^{-1} t^{-1} \I_{L_{\mathrm{ini}}R_{\mathrm{ini}}}), \label{eq:pA_1} \\
    p(\B | t,\beta) & = & \prod^J_{j=1}\N(\boldsymbol{b}_j\mid\0,\beta^{-1} t^{-1} \I_{L_{\mathrm{ini}}R_{\mathrm{ini}}}), \label{eq:pB_1} \\
    p(\C | t,\beta) & = & \prod^K_{k=1}\N(\boldsymbol{c}_k\mid\0,\beta^{-1} t^{-1} \I_{R_{\mathrm{ini}}}), \label{eq:pC_1} 
\end{eqnarray}
where $t$ is now a deterministic parameter, intended to play the role of the regularization parameter in the associated deterministic regularization-based problem.  The corresponding graphical model is given in Fig.~\ref{fig:graphs}-I. 
It is obviously a simplified, single-level version of the 3-level hierarchical model introduced in Section~\ref{sec:Bayes}.

To perform point estimation of $\A,\B$ and $\C$, the corresponding MAP estimator is derived next. The joint posterior pdf of $\A,\B,\C$ can be expressed as follows:  
\begin{equation}
p(\A,\B,\C|\bc{Y},\beta) \propto p(\bc{Y}|\A,\B,\C,\beta)p(\A,\B,\C|\beta).
\label{eq:jpo_1a}
\end{equation}
Due to the Gaussianity of the noise, the likelihood function $p(\bc{Y}|\A,\B,\C,\beta)$ can be  written from~\eqref{eq:Y=X+N} as 
\begin{equation}
p(\bc{Y}|\A,\B,\C,\beta)\propto \exp\left(-\frac{\beta}{2}\left\|\bc{Y}-\sum_{r=1}^{R_{\mathrm{ini}}}\mb{A}_{r}\mb{B}_{r}^{\T}\circ \mb{c}_{r}\right\|_{\F}^2\right).   
\label{eq:lfun}
\end{equation}
In addition, from~\eqref{eq:pA_1}, \eqref{eq:pB_1},  and~\eqref{eq:pC_1}, the prior distribution of $\A,\B$ and $\C$ takes the following form
\begin{align}
 p(\A,\B,\C|\beta) &\propto \exp \left[-\frac{\beta t}{2} \sum_{r=1}^{R_{\mathrm{ini}}}\sum_{l=1}^{L_{\mathrm{ini}}}(\|\mb{a}_{r,l}\|_{2}^{2}+\|\mb{b}_{r,l}\|_{2}^{2})\right]  \nonumber \\
 & \times \exp \left( -\frac{\beta t}{2} \sum_{r=1}^{R_{\mathrm{ini}}}\|\mb{c}_r\|^{2}_2\right).
 \label{eq:jprior_1b}
\end{align}
Combining~\eqref{eq:lfun} with~\eqref{eq:jprior_1b} and taking the logarithm of their product we end up with the following MAP-type optimization problem:
\begin{equation}
\min_{\A,\B,\C} \left\|\bc{Y}-\sum_{r=1}^{R_{\mathrm{ini}}}\mb{A}_{r}\mb{B}_{r}^{\T}\circ \mb{c}_{r}\right\|_{\F}^2 +
 t\left(\|\mb{A}\|_{\F}^2+\|\mb{B}\|_{\F}^2+\|\mb{C}\|_{\F}^2\right).
\label{eq:map_model1}
\end{equation} 
This implies that the regularizer induced by the single-level Gaussian priors favors smooth solutions in terms of the latent factors. This is deduced from the fact that $\A,\B$ and $\C$ can be updated using an alternating minimization strategy which gives rise to ridge regression-type subproblems (as it is done in, e.g., \cite{qxzzt17}). In the light of this feature, no distinction between the columns of each of the factor matrices is being made and hence Model~I is expected to have a weaker rank revelation effect than the one of the main model introduced in Section \ref{sec:Bayes}. 

\paragraph{Model~II}
With the latter observation in mind, we now introduce the second alternative Bayesian model whose graphical model is depicted in Fig.~\ref{fig:graphs}-II.
Model~II places heavy-tailed multi-parameter Laplace priors, known for their sparsity-inducing effect, over the columns of the factor matrices. This is implemented with the aid of a three-level hierarchy of priors. At the first level, Gaussian distributions are assigned to the factors, namely,
\begin{eqnarray}
    p(\A | \mb{t}_A,\beta) & = & \prod^I_{i=1}\N(\boldsymbol{a}_i\mid\0,\beta^{-1}\mb{T}_A^{-1}), \label{eq:pA2} \\
    p(\B | \mb{t}_B,\beta) & = & \prod^J_{j=1}\N(\boldsymbol{b}_j\mid\0,\beta^{-1}\mb{T}_B^{-1}), \label{eq:pB2} \\
    p(\C | \zet,\beta) & = & \prod^K_{k=1}\N(\boldsymbol{c}_k \mid\0,\beta^{-1}\Z^{-1}), \label{eq:pC2}
    \end{eqnarray}
where $\mb{T}_A=\mathrm{diag}({\mb{t}_A})$ and $\mb{T}_B =\mathrm{diag}({\mb{t}_B})$ with $\mb{t}_A, \mb{t}_B \in \Rb^{L_{\mathrm{ini}}R_{\mathrm{ini}}\times 1}$ and $\Z=\mathrm{diag}(\zet)$, $\zet\in\Rb^{R_{\mathrm{ini}}\times 1}$. Note that the key difference of this model with the one introduced in Section~\ref{sec:Bayes} and depicted in Fig.~\ref{fig:graph} is the use of different variables ${\mb{t}_A},{\mb{t}_B}$ for enforcing column sparsity on $\A$ and $\B$. This is in contrast to the ``coupling" of $\mb{A}_r,\mb{B}_r$ effected in BTD-HIRLS and the model of Fig.~\ref{fig:graph}. Moreover, the parameters $\zet$ are now involved only in the prior of $\C$, which again ``decouples" the third mode factor from the rest. 

At the second level of the hierarchy, IG priors are placed over $\mb{t}_A, \mb{t}_B$ and $\zet$, namely
\begin{eqnarray}
    p(\mb{t}_A) & = & \prod_{r=1}^{R_{\mathrm{ini}}}\prod_{l=1}^{L_{\mathrm{ini}}} \IG\left (t_{A;r,l}\left | \frac{I+1}{2},\frac{\delta_{A;r,l}}{2}\right.\right), \\
    p(\mb{t}_B) & = & \prod_{r=1}^{R_{\mathrm{ini}}}\prod_{l=1}^{L_{\mathrm{ini}}} \IG\left (t_{B;r,l}\left | \frac{J+1}{2},\frac{\delta_{B;r,l}}{2}\right.\right), \\
    p(\zet) & = & \prod^{R_{\mathrm{ini}}}_{r=1}\IG\left (\zeta_r \left |\frac{K+1}{2},\frac{\rho_r}{2}\right.\right),
\end{eqnarray}
where $\delta_{A;r,l}, \delta_{B;r,l}$ and $\rho_r$ are the scale parameters of the distributions over $t_{A;r,l}, t_{B;r,l}$ and $\zeta_r$, respectively.
The third level involves Gamma priors over these variables, namely, 
\begin{align}
p(\delta_{A;r,l}) = \G(\delta_{A;r,l}\mid \psi_A,\tau_A),\\
      p(\rho_{A,r}) = \G(\rho_{A,r} \mid \mu_A,\nu_A),
\end{align}
and similarly for $\delta_{B;r,l},\rho_{B,r}$.

This ``decoupling" approach allows us to derive the MAP estimator for $\A,\B,\C$ and thus gain a deeper insight as to the regularization effect induced by the model. As explained earlier, a MAP-type problem cannot be derived for the main model presented in Section~\ref{sec:Bayes}  due to the interrelation among different variables. Yet, the increased complexity of that model better captures the structure of BTD, as it is also empirically demonstrated in the experimental results. For Model~II, the joint prior pdf of $\A,\B,\C$ can be computed from the following multiple integral 
\begin{eqnarray}
\lefteqn{p(\A,\B,\C|\beta,\delt_{A},\delt_{B},\ro)=} \nonumber \\
& & \!\!\!\!\!\!\!\int p(\A,\B,\C|\beta,\mb{t}_{A},\mb{t}_{B},\zet)\times \nonumber \\
& & p(\mb{t}_{A}|\delt_{A})p(\mb{t}_{B}|\delt_{B})p(\zet|\ro)\mathrm{d}\mb{t}_{A}\mathrm{d}\mb{t}_{B}\mathrm{d}\zet,
\label{eq:jprior}
\end{eqnarray}
where
\begin{align}
p(\A,\B,\C|\beta,\mb{t}_{A},\mb{t}_{B},\zet)&=\prod_{r=1}^{R_{\mathrm{ini}}}\prod_{l=1}^{L_{\mathrm{ini}}}p(\mb{a}_{r,l},\mb{b}_{r,l}|\beta,t_{r,l})  \nonumber \\
&\times \prod_{r=1}^{R_{\mathrm{ini}}}p(\mb{c}_r|\beta,\zeta_r).
\label{eq:jprior1}
\end{align}
After substituting~\eqref{eq:jprior1} to~\eqref{eq:jprior} we get the expression for the joint prior distribution shown in~\eqref{eq:jprior2} at the top of the next page.
\begin{figure*}
\begin{align}
p(\A,\B,\C|\beta,\delt_{A},\delt_{B},\ro) &=  \prod_{r=1}^{R_{\mathrm{ini}}}\prod_{l=1}^{L_{\mathrm{ini}}}\int_{0}^{\infty}p(\mb{a}_{r,l},\mb{b}_{r,l}|\beta,t_{A;r,l},t_{B;r,l})
p(t_{A;r,l}|\delta_{A;r,l} )p(t_{B;r,l}|\delta_{B;r,l})\mathrm{d}t_{A;r,l}\mathrm{d}t_{B;r,l} \nonumber \\
&\times\prod_{r=1}^{R_{\mathrm{ini}}}\int_{0}^{\infty}p(\mb{c}_r|\beta,\zeta_r)p(\zeta_r|\rho_r)\mathrm{d}\zeta_r.
\label{eq:jprior2}
\end{align}
\end{figure*}
The integrals in~\eqref{eq:jprior2} can be computed by working as in~\cite[Appendix B]{grtk17} whereby the joint prior pdf of $\A,\B,\C$ results as
\begin{align}
 &p(\A,\B,\C|\beta,\delt,\ro) \propto \nonumber \\
 &\exp \left[ -\beta^{\frac{1}{2}}\sum_{r=1}^{R_{\mathrm{ini}}}\sum_{l=1}^{L_{\mathrm{ini}}}(\delta_{A;r,l}^{\frac{1}{2}}\|\mb{a}_{r,l}\|_{2}+\delta_{B;r,l}^{\frac{1}{2}}\|\mb{b}_{r,l}\|_{2})\right]  \nonumber \\
 & \times \exp \left( -\beta^{\frac{1}{2}}\sum_{r=1}^{R_{\mathrm{ini}}}\rho_{r}^{\frac{1}{2}}\|\mb{c}_r\|_2\right), 
\label{eq:jprior3}
\end{align}
which is a heavy-tailed multi-parameter multivariate Laplace distribution defined on the columns of $\mb{A},\mb{B}$ and $\mb{C}$. From~\eqref{eq:lfun} and~\eqref{eq:jprior3}, the MAP estimator of Model~II is obtained from the solution of the following minimization problem
\begin{align}
&\min_{\A,\B,\C} \frac{\beta}{2}\left\|\bc{Y}-\sum_{r=1}^{R_{\mathrm{ini}}}\mb{A}_{r}\mb{B}_{r}^{\T}\circ \mb{c}_{r}\right\|_{\F}^2 + \nonumber \\
&\beta^{\frac{1}{2}}\left(\sum_{r=1}^{R_{\mathrm{ini}}}\sum_{l=1}^{L_{\mathrm{ini}}}(\delta_{A;r,l}^{\frac{1}{2}}\|\mb{a}_{r,l}\|_{2}+\delta_{B;r,l}^{\frac{1}{2}}\|\mb{b}_{r,l}\|_{2}) + \sum_{r=1}^{R_{\mathrm{ini}}}\rho_{r}^{\frac{1}{2}}\|\mb{c}_r\|_2\right).
\label{eq:map_model2}
\end{align}

\noindent
\emph{Remark:} It should be noted that~\eqref{eq:map_model2} bears a close resemblance to the deterministic criterion proposed in~\cite{gofzc20} and can thus be seen to offer a Bayesian interpretation thereof and suggest the corresponding Bayesian inference method as a probabilistic counterpart of the \emph{Alternating Group Lasso (AGL)} algorithm developed in~\cite{gofzc20}. Given the correspondence of the main model, in Section~\ref{sec:Bayes}, with BTD-HIRLS, the comparison of these Bayesian methods presented in the next section complements in a way the comparative study of BTD-HIRLS and AGL previously reported~\cite{rkg21}.
 
%See in~\cite{pg08} how $\lambda$ is updated in each iteration, based on the posterior means of the coefficient variances, computed on the basis of the $\lambda$ posterior. With small powers, $\lambda$ is increased to force sparsity. 

\section{Simulation Results}
\label{sec:sims}

In this section, we evaluate the effectiveness of the proposed algorithm in Table~\ref{tab:BBTD} in selecting and computing the appropriate BTD model for a given tensor, via simulations with both synthetic and real data. Its deterministic counterpart from~\cite{rkg21} and the corresponding Bayesian inference methods resulting from Models~I, II and referred to henceforth as BBTD~model~I and BBTD~model~II are included, for comparison purposes. In the last experiment, and in a hyperspectral image denoising problem, BBTD is also compared with the Bayesian CPD method that emanates from Model~II as a special case with $L_r=1$, $r=1,2,\ldots,R$. 

\subsection{Synthetic data experiments}

In this part, we first test the BBTD algorithm of Table~\ref{tab:BBTD} in comparison to its alternatives. We also demonstrate its robustness to initialization and compare its ability to recover the correct ranks of the BTD model against BTD-HIRLS. The adopted figure of merit is the \emph{Normalized Mean Squared Error (NMSE)} over block terms, defined as $\mathrm{NMSE} =\sum^R_{r=1}\frac{\|\A_r\B_r^{\T} \circ \c_r - \hat{\A}_r\hat{\B}_r^{\T}\circ\hat{\c}_r\|^2_{\F}}{\|\A_r\B_r^{\T} \circ \c_r\|^2_{\F}}$. As in~\cite{rkg21}, the Hungarian algorithm is employed to match the $\hat{R}$ estimated non-zero block terms with the true ones. 

\paragraph{Performance comparison between the main model and Models~I and~II}

\begin{figure}
    \centering
    \includegraphics[width=0.4\textwidth]{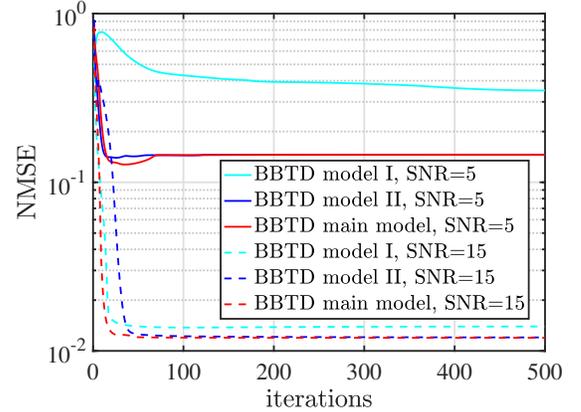}
    \caption{NMSE vs. iterations for the proposed BBTD algorithm (`BBTD main model') and its variants (`BBTD model~I', `BBTD model~II') at two SNR values.}
    \label{fig:mod_comp}
\end{figure}

In this experiment, we generate $18\times 18\times 10$ tensors $\bc{Y}$ as in~(\ref{eq:Y=X+N}), with $R=3$ and the $L_r$s set as $L_1=8,L_2=6$ and $L_3=4$. The entries of $\A_r,\B_r$ and $\C$ are i.i.d., sampled from the standard Gaussian distribution. The noise power is set so as to result in a signal-to-noise ratio $\mathrm{SNR}=10\log_{10} \|\bc{X}\|_{\F}^2/(\sigma^2\|\bc{N}\|_{\F}^2)$ of~5 and 15~dB. Both $R$ and all $L_r$s are overestimated as $R_{\mathrm{ini}}=L_{\mathrm{ini}}=10$. Fig.~\ref{fig:mod_comp} illustrates the best run in terms of the NMSE, obtained out of~10 random initializations of the algorithms. 
As it can be observed, the main model described in Section~\ref{sec:Bayes} performs comparably to Model~II, which can be viewed as a relaxed version thereof. Notably, the algorithm of Table~\ref{tab:BBTD} converges somewhat faster. It is worth noting that the algorithm associated to Model~I exhibits a poorer performance -- especially at low SNR -- due to its inaptitude in dealing effectively with the over-parameterized regime when it comes to the BTD ranks. As opposed to Model~I, both the main model and Model~II, which use heavy-tailed priors on the latent BTD factors, show their efficacy in addressing the challenges incurred by the unawareness of BTD ranks and successfully model the tensors at both SNR values examined. 

In the following, we focus on the BBTD algorithm of Table~\ref{tab:BBTD}, which better captures the structure of the BTD model. At the same time it makes use of fewer latent variables at the second and third level of hierarchy than those in Model~II and hence we consider it as a more compact version of the latter.

\paragraph{Robustness to initialization} 

\begin{figure}
    \centering
    \includegraphics[width=0.45\textwidth]{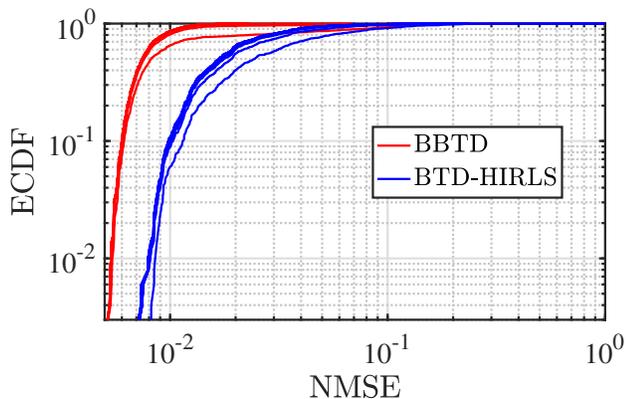}
    \caption{Empirical Cumulative Distribution Function (ECDF) of NMSE obtained by BBTD and BTD-HIRLS from 500~independent runs. The $i$th curve from bottom to top corresponds to the result of selecting the best out of $i = 1, 2,\ldots, 12$ different initializations. SNR=15~dB.}
    \label{fig:ecdf}
\end{figure}

In an effort to see how robust the proposed BBTD algorithm is to initialization, we set SNR=15~dB and generate tensors as previously. We run 500~realizations of the experiment. For each, we apply the proposed BBTD and the BTD-HIRLS algorithms with 12~different random initializations. Fig.~\ref{fig:ecdf} shows the Empirical Cumulative Distribution Function (ECDF) of the obtained NMSE, where the $i$th curve from bottom to top corresponds to selecting the best out of $i$ initializations, for $i = 1, 2, \ldots, 12$. 
It can be observed that  BBTD is rather insensitive to initialization. Surprisingly, its performance is affected by random initialization even less than BTD-HIRLS, whose robustness has also been verified~\cite{rkg21}. We thus have empirical evidence that only a small number of initializations suffices to estimate an accurate BTD model with the proposed BBTD algorithm. 

\paragraph{Rank recovery} 

\begin{figure}
    \centering
    \begin{tabular}{c c} \includegraphics[width=0.22\textwidth]{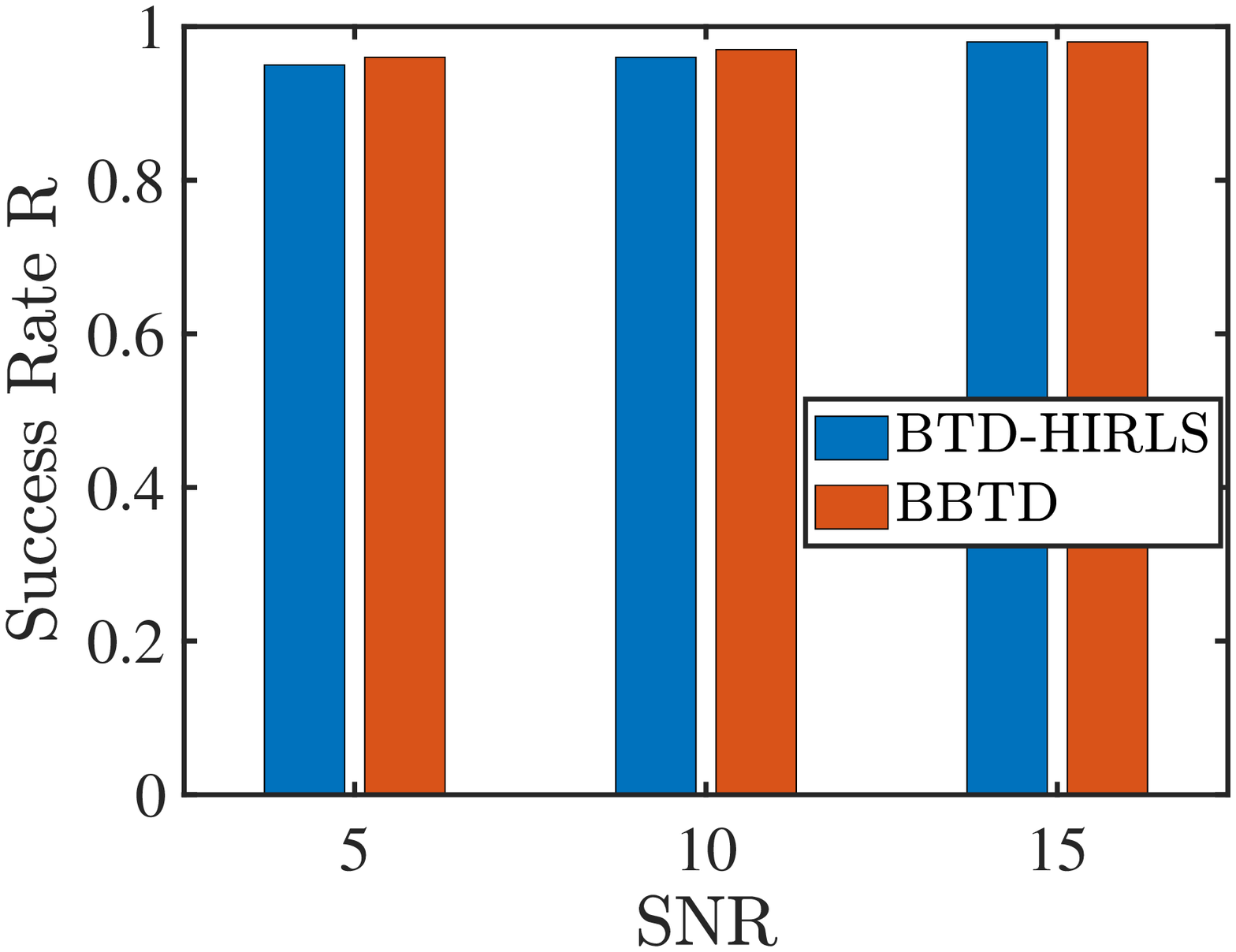} &\includegraphics[width=0.22\textwidth]{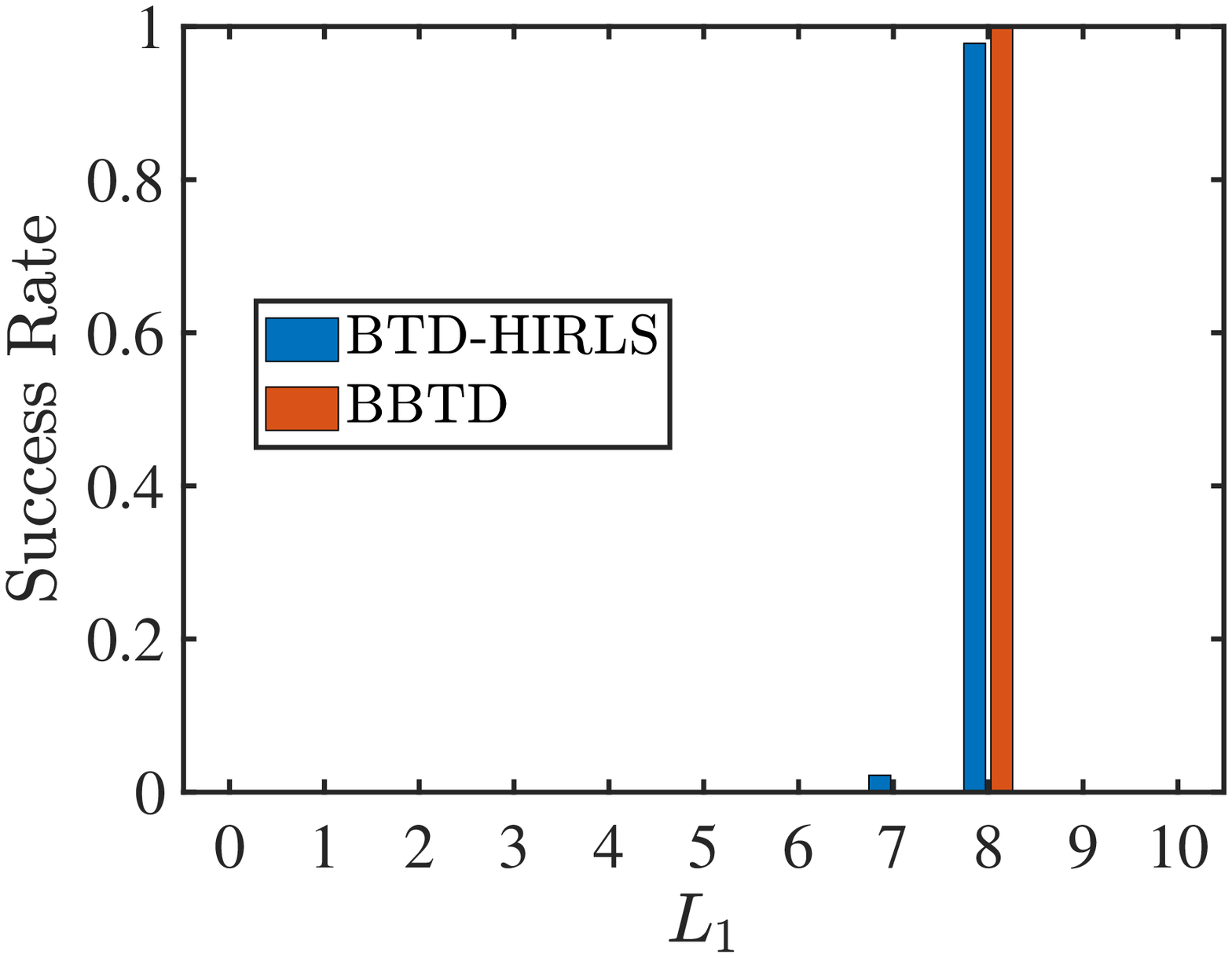}
      \\
    (a) & (b) \\
    \includegraphics[width=0.22\textwidth]{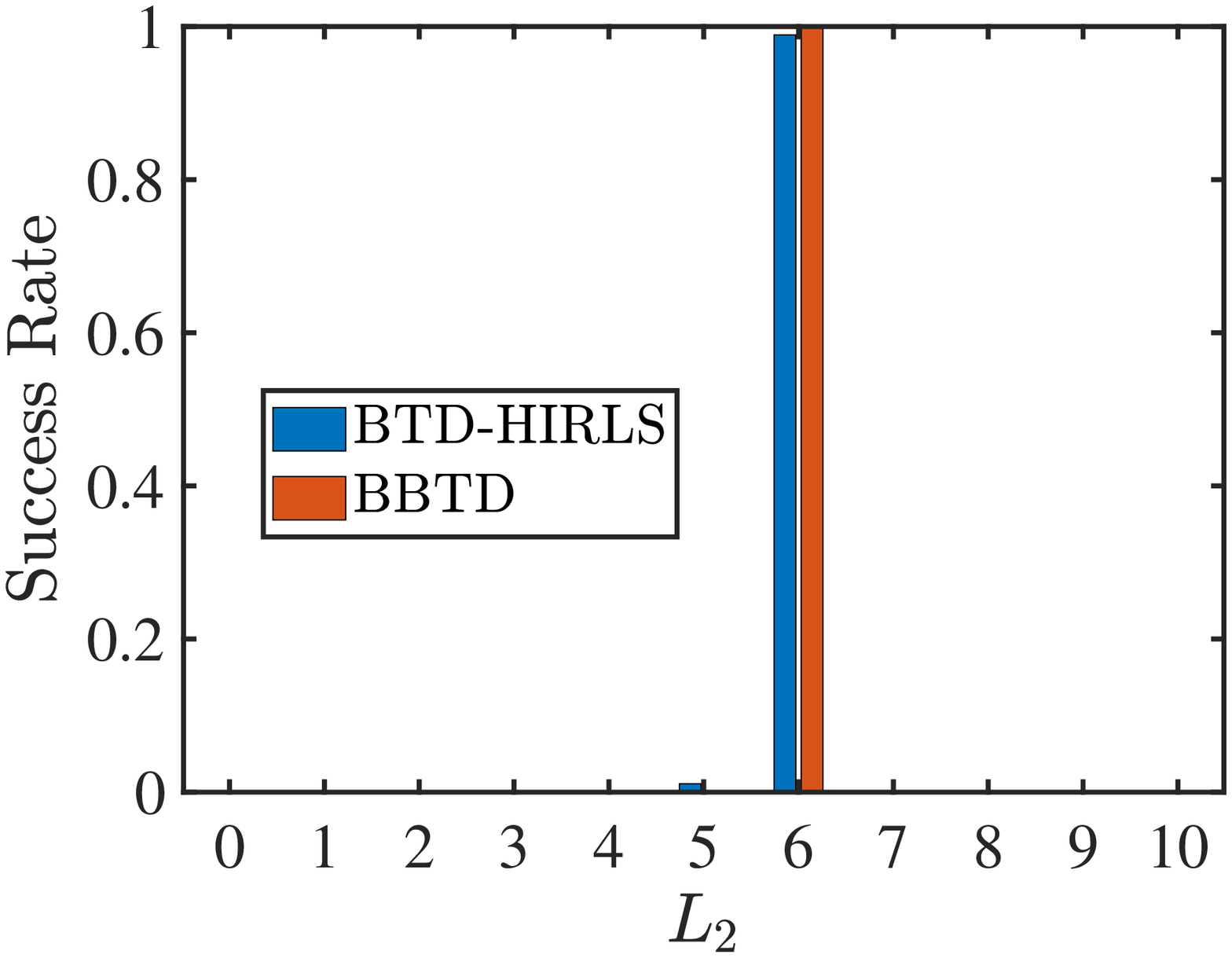} &\includegraphics[width=0.22\textwidth]{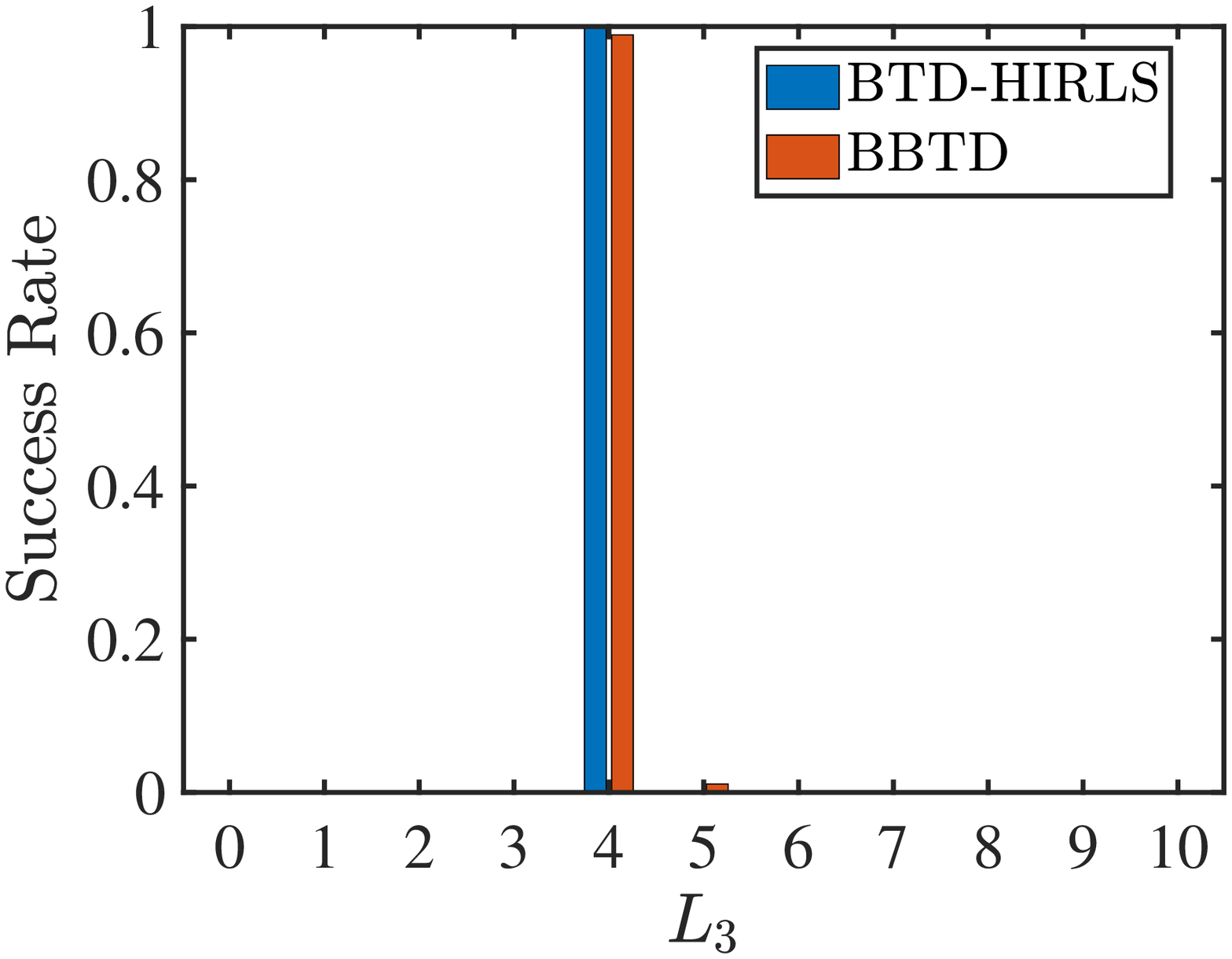}
      \\
    (c) & (d) \\
    \includegraphics[width=0.22\textwidth]{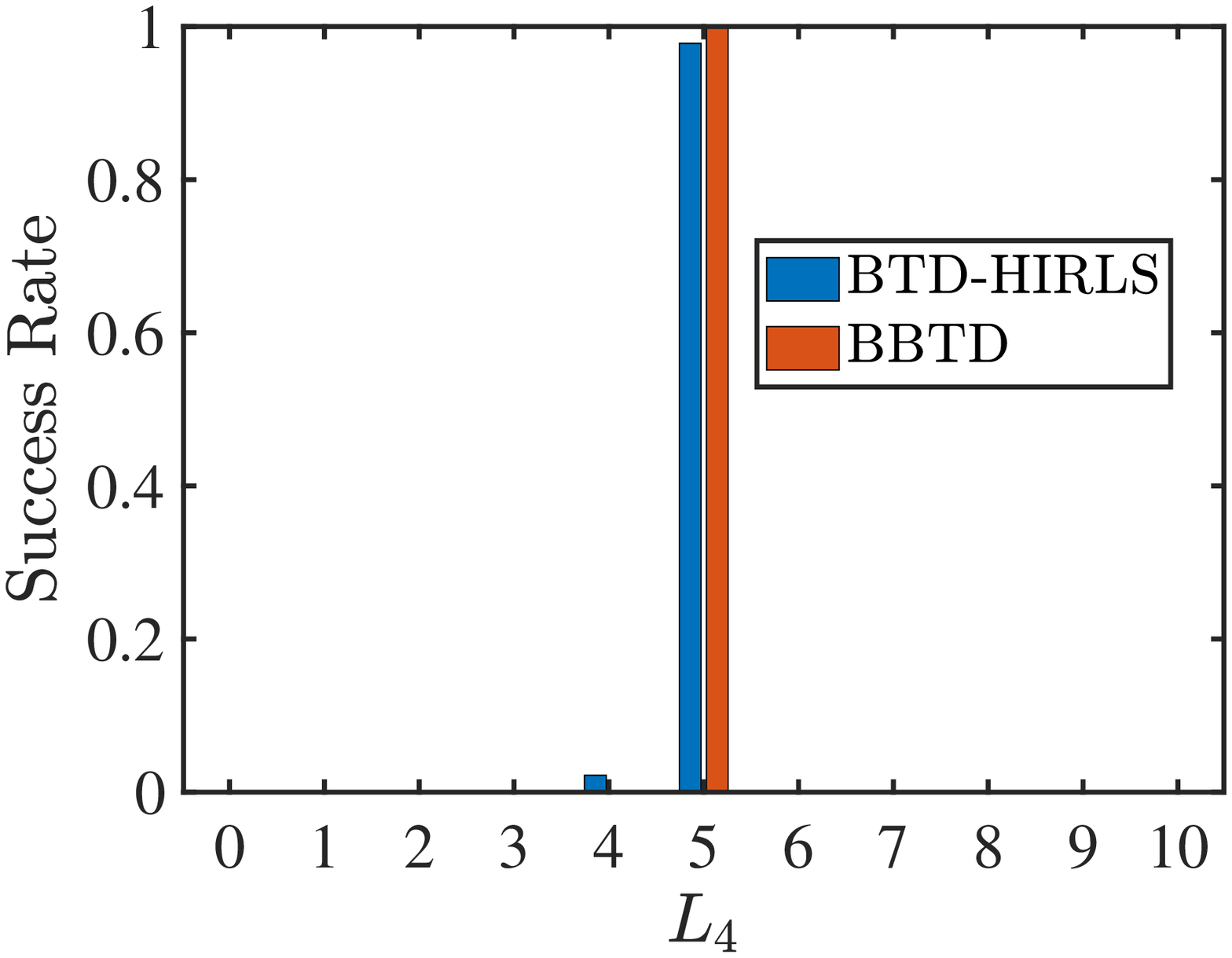} &\includegraphics[width=0.22\textwidth]{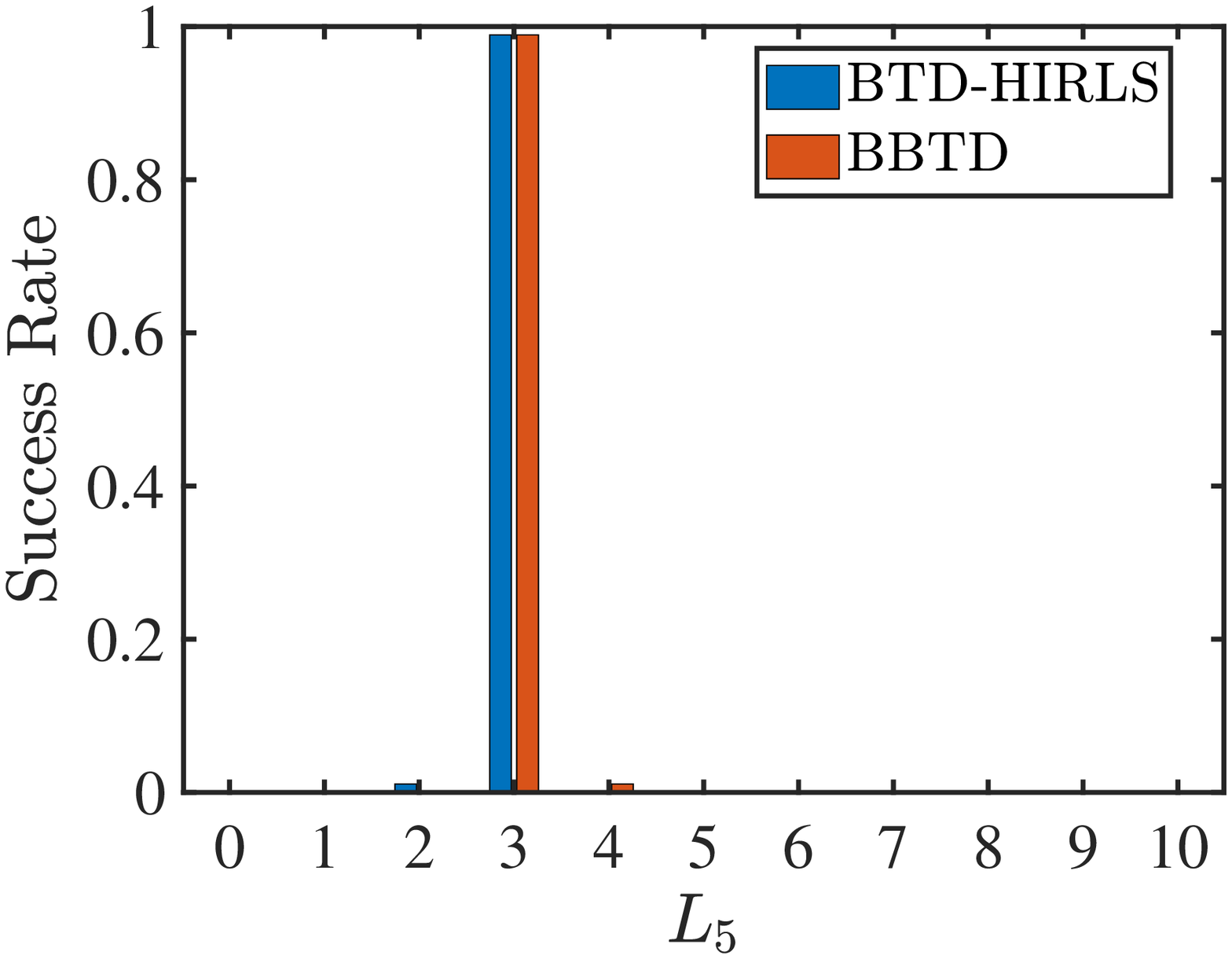}
      \\
    (e) & (f)
    \end{tabular}
    \caption{Success rates of recovering (a) $R$ and (b)--(f) $L_r$s for SNR=10~dB with the aid of the BBTD and BTD-HIRLS algorithms. Scenario~A: $\min(I,J)>\sum_{r=1}^R L_r$.}
    \label{fig:succ_rates_A}
\end{figure}

Here we use the same generative model described above for building  data tensors $\bc{Y}$ of dimensions $30\times 30\times 30$. Our objective is to assess the ability of  BBTD to select the correct BTD model. For comparison purposes, we also employ the BTD-HIRLS algorithm, which has demonstrated high model selection ability in~\cite{rkg21}. Two different scenarios are considered, that differ in the validity of the well-known sufficient BTD uniqueness condition of having full column rank $\mb{A},\mb{B}$ matrices and a $\mb{C}$ matrix with non-collinear columns~\cite{ldl08b}.\\
\emph{Scenario A:} In this scenario, we set $R=5$ and the true $L_r$s are set to $L_1=8, L_{2}=6, L_{3}=4, L_{4}=5$ and $L_{5}=3$. This setting is favorable w.r.t. the above condition since $\min(I,J) > \sum^{R}_{r=1}L_{r}$. Fig.~\ref{fig:succ_rates_A}(a) shows the success rates of the recovery of $R$ for SNR equal to 5, 10, and 15~dB. 
BBTD performs slightly better than BTD-HIRLS at~5 and~10~dB and has similar performance at~15~dB. Note that BTD-HIRLS required its regularization parameter to be finely tuned, whereas this is automatically performed in the BBTD algorithm, in a data-driven way. In Figs.~\ref{fig:succ_rates_A}(b)--(f), and restricting attention to those realizations where both algorithms have succeeded in recovering the true value of $R$, the success rates of recovering the block ranks $L_r$ are depicted, at an SNR value of~10~dB. Observe that there is an almost 100\% success for all $R$ terms. These results  provide empirical evidence of the competence of BBTD in this challenging, yet critical, task of inferring the correct model structure.

\begin{figure}
    \centering
    \begin{tabular}{c c} \includegraphics[width=0.22\textwidth]{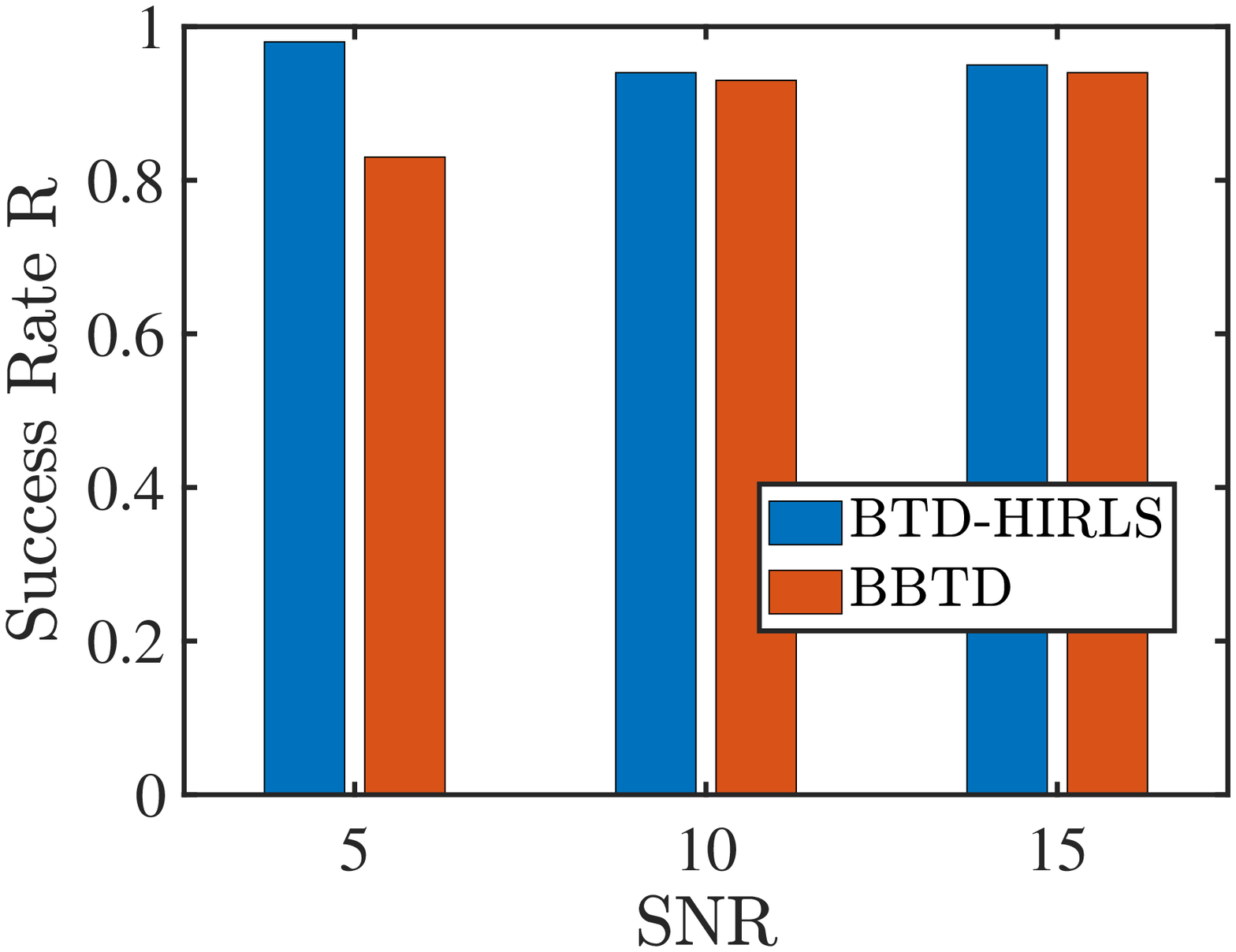} &\includegraphics[width=0.22\textwidth]{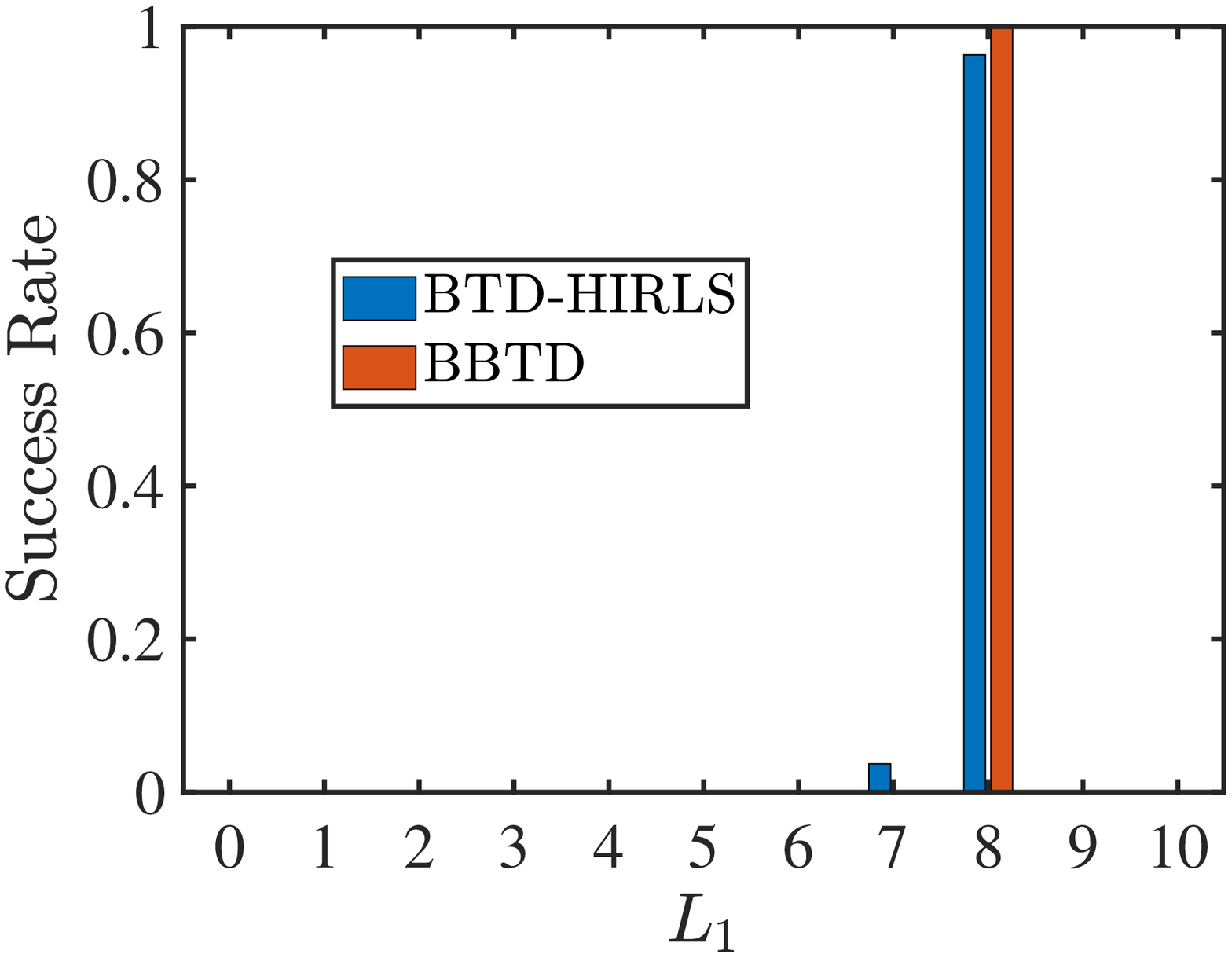}
      \\
    (a) & (b) \\
    \includegraphics[width=0.22\textwidth]{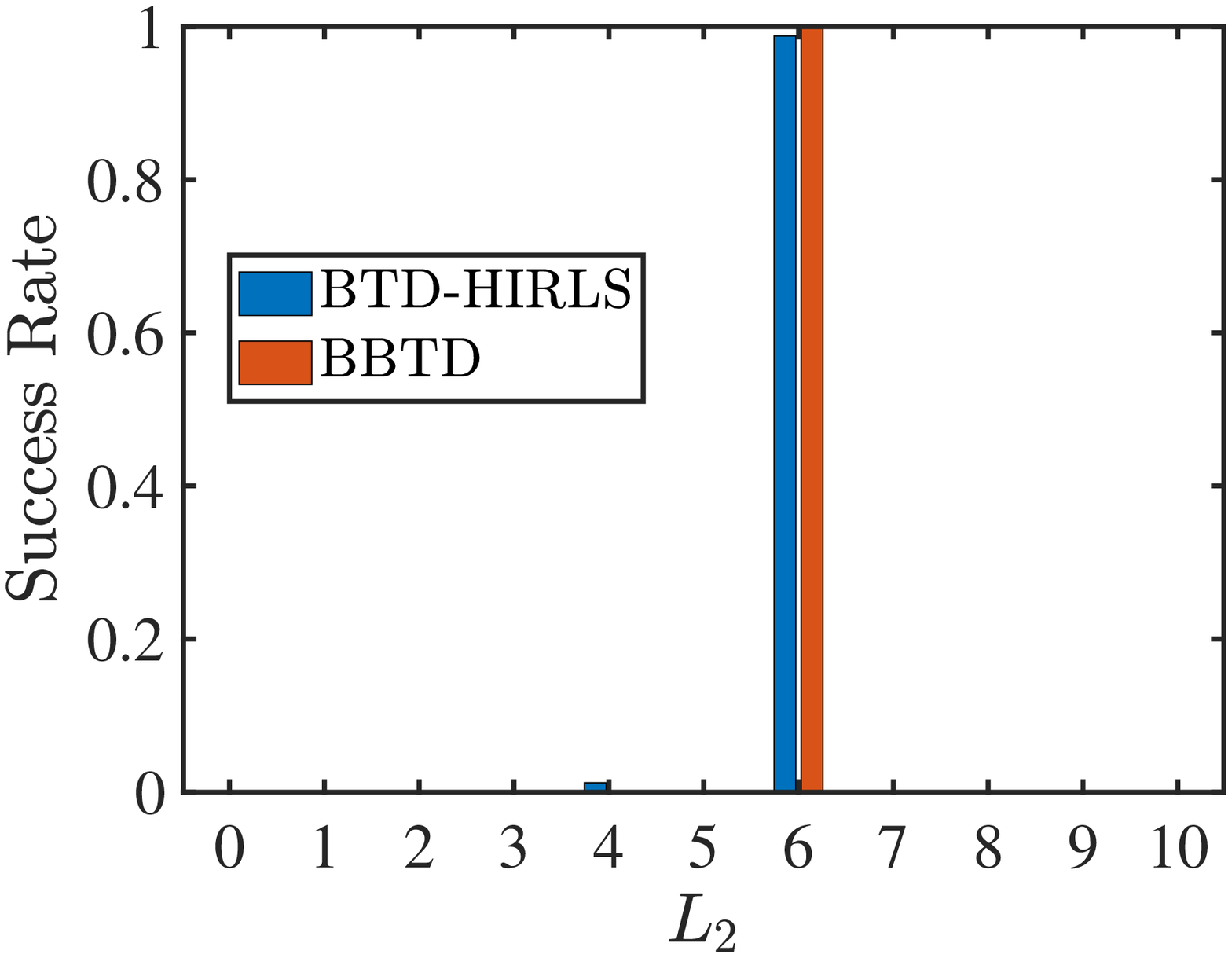} &\includegraphics[width=0.22\textwidth]{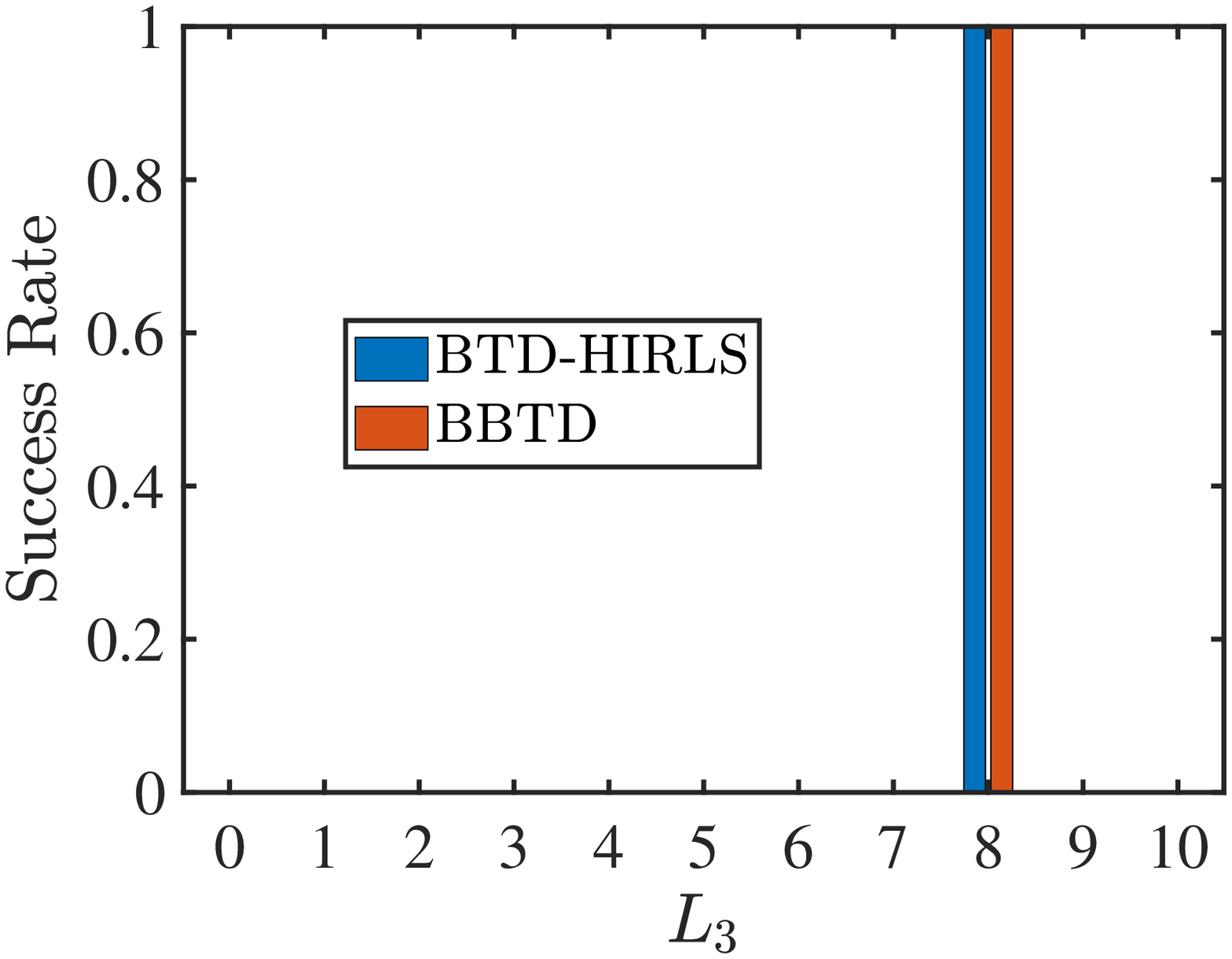}
      \\
    (c) & (d) \\
    \includegraphics[width=0.22\textwidth]{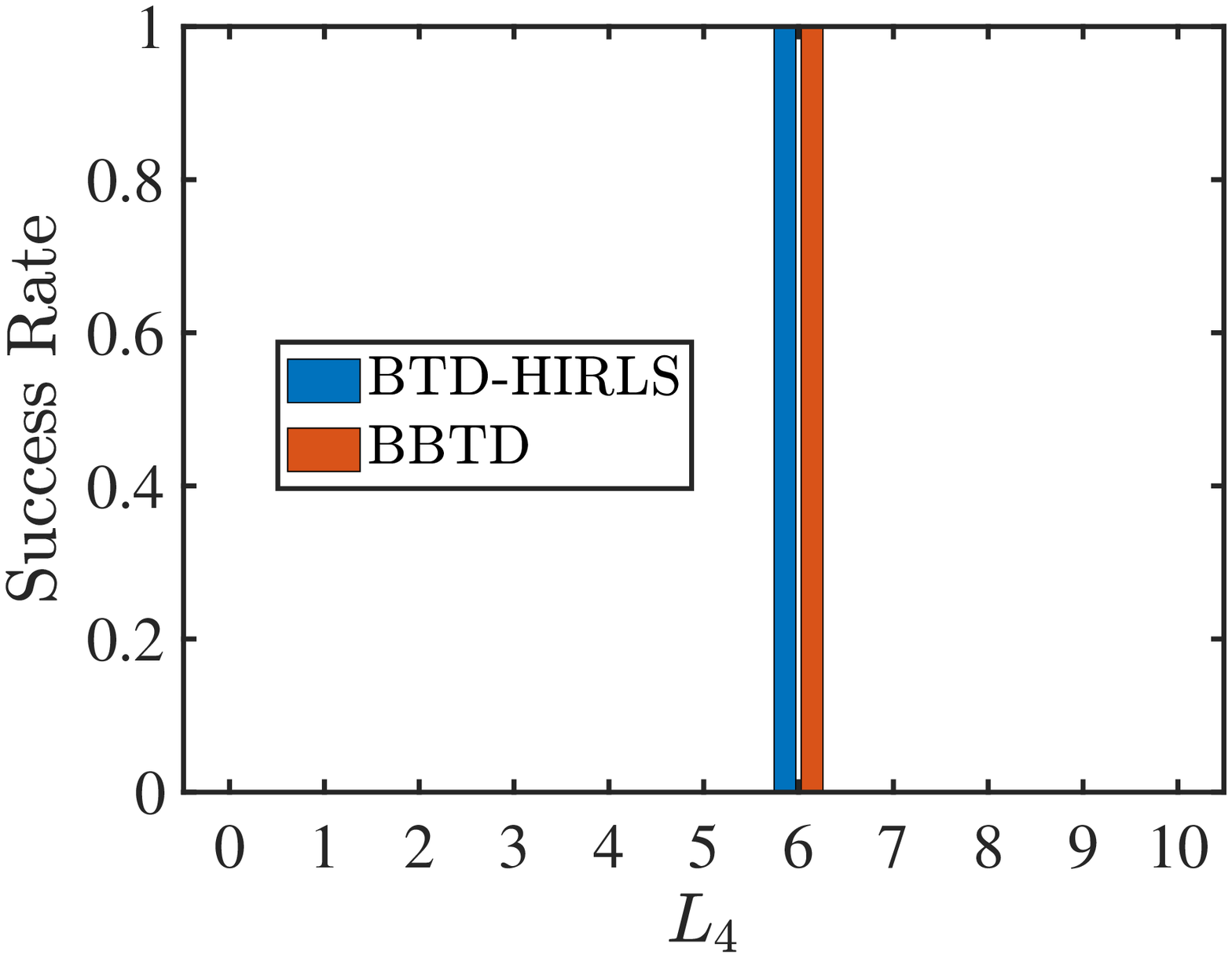} &\includegraphics[width=0.22\textwidth]{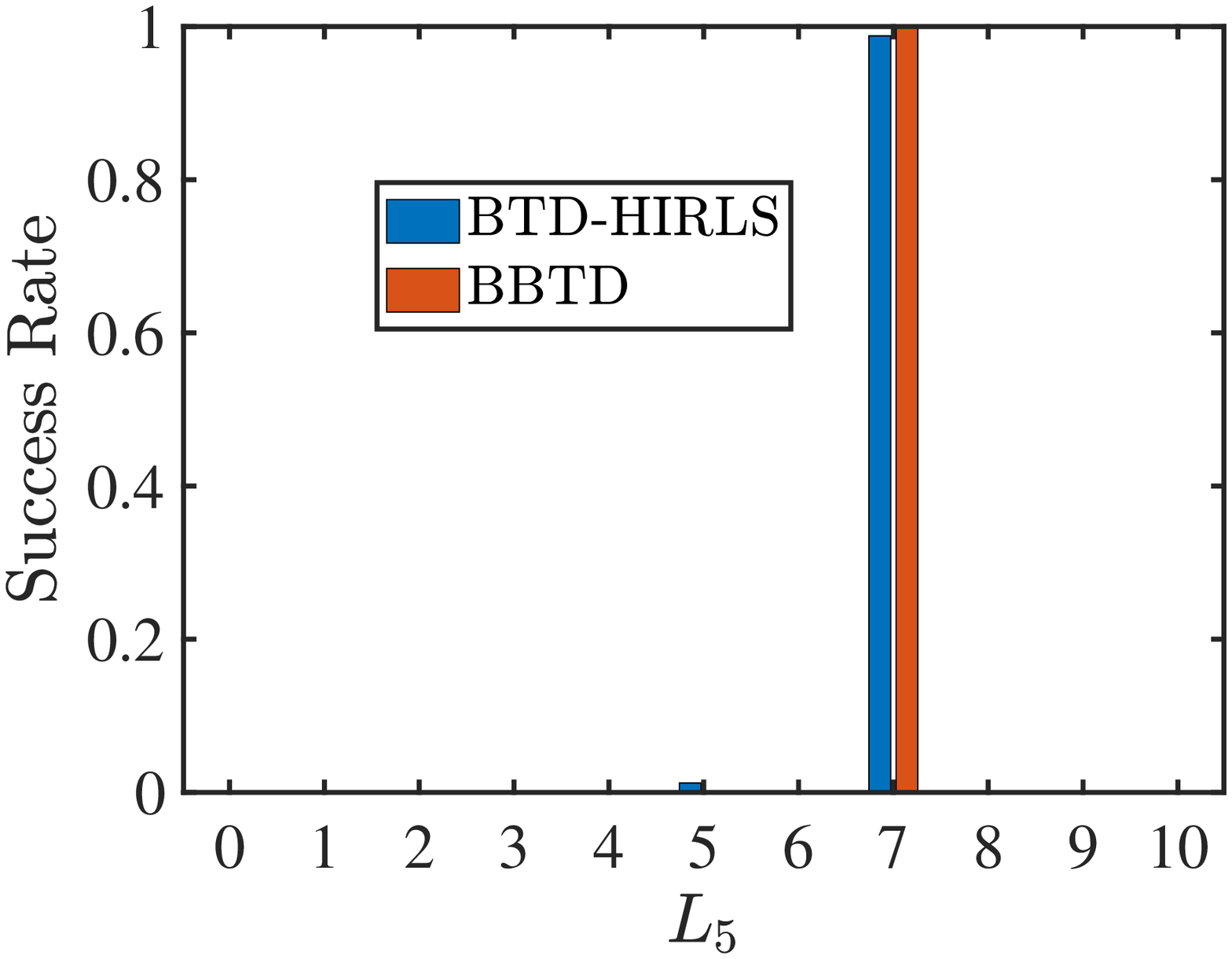}
      \\
    (e) & (f)
    \end{tabular}
    \caption{Success rates of recovering (a) $R$ and (b)--(f) $L_r$s for SNR=10~dB with the aid of the BBTD and BTD-HIRLS algorithms. Scenario~B: $\min(I,J)<\sum_{r=1}^R L_r$.}
    \label{fig:succ_rates_B}
\end{figure}

\emph{Scenario B:}  We now choose the following values for the block ranks, $L_1=8,L_{2}=6,L_{3}=8,L_{4}=6$ and $L_{5}=7$, for which $\min(I,J)<\sum^{R}_{r=1}L_{r}$ and hence the sufficient uniqueness condition is no longer satisfied. This experimental setting is therefore considered to be even more  challenging than Scenario~A. As shown in Fig.~\ref{fig:succ_rates_B}(a), BBTD performs comparably to the BTD-HIRLS in estimating $R$, with the latter being somewhat better at SNR=5~dB. 
It should, however, be reminded that this is the result of tuning the regularization parameter, a task that can be far from being easy in real-world applications. Moreover, a behavior similar to that in Scenario~A is observed when it comes to the success rates of recovering the $L_r$s at SNR=10~dB; see Figs.~\ref{fig:succ_rates_B}(b)--(f). BBTD is again slightly superior to BTD-HIRLS in carrying out this intricate task while enjoying the advantage of being completely automatic.

\subsection{Real data experiment: Hyperspectral image denoising}
\label{sec:real}

Hyperspectral imagery (HSI) can be represented with the aid of 3-way tensors whose first two modes correspond to the spatial domain and the third one to the spectral domain. It is known that there is inherent correlation in both domains, which explains the fact that low-rank matrix and tensor  representations have been widely adopted for numerous HSI processing tasks such as unmixing~\cite{gtrk16,qxzzt17} and  restoration~\cite{xzq19}. It should be emphasized that the very nature of HSI, accurately described by a linear mixing model~\cite{qxzzt17}, points to BTD as the most suitable choice of a decomposition model as compared to classical CPD. Indeed, the model structure and parameters are in a direct correspondence with the HSI constituents: the $R$ matrices $\mb{E}_r$ can be interpreted as the abundance maps while $\mb{C}$ contains the endmember spectral signatures in its columns. 

As an example of the application of our method in this context, we consider the problem of denoising a hyperspectral image, and compare with the results of BTD-HIRLS and a Bayesian CPD method resulting from BTD~model~II as a special case and referred to here as BCPD. We generate a noisy version of the Washington DC Mall AVIRIS image by adding i.i.d. Gaussian noise and choosing its power so as to get SNR=5~dB. The HSI volume is captured at $K=191$ contiguous spectral bands in the~0.4 to~2.4$\mu$m region of the visible infrared spectrum~\cite{grk19a}. The size of the image at each spectral band is $150\times150$ pixels and hence the HSI cube can be seen as a $150\times150\times 191$ tensor. Our objective is to suppress the noise by fitting a decomposition model to this tensor. Of course, the correct $R$ and $L_r$s must also be estimated. To this end, they are overestimated as $R_{\mathrm{ini}}=50$ and $L_{\mathrm{ini}}=10$. Finally, we initialize the tensor rank of BCPD to $R_{\mathrm{BCPD,ini}} = R_{\mathrm{ini}}L_{\mathrm{ini}} = 500$.

\begin{figure}
    \begin{center}
    \includegraphics[width=0.47\textwidth]{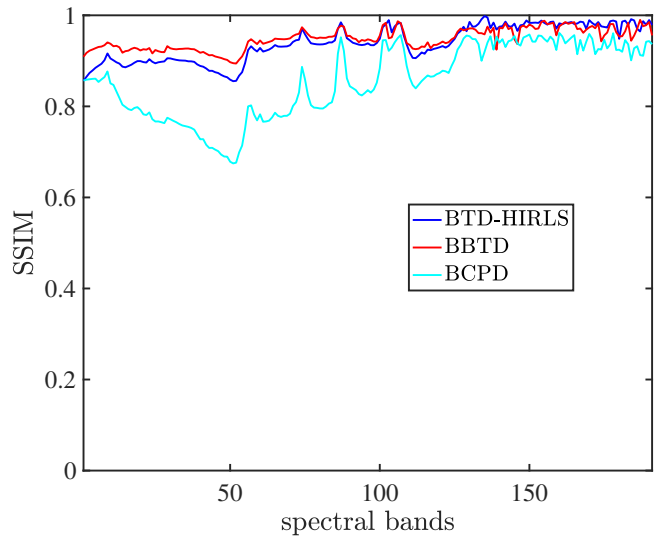}
    \caption{SSIM of the hyperspectral images recovered by BBTD, BTD-HIRLS and BCPD.}
            \label{fig:ssim}
    \end{center}
\end{figure}

We compare the performance of the three methods both visually and in terms of the \emph{Structural Similarity Index Measure (SSIM)}, a popular perceptual metric of the degradation of an image as perceived change in structural information. SSIM is defined for two image windows $x,y$ as $\mathrm{SSIM}(x,y)=\frac{(2\mu_{x}\mu_{y}+c_{1})(2\sigma_{xy}+c_2)}{(\mu^{2}_{x}+\mu^{2}_{y}+c_{1})(\sigma^{2}_{x}+\sigma^{2}_{y}+c_{2})}$, where $\mu_{x},\mu_{y},\sigma^{2}_{x},\sigma^{2}_{y}$ are their mean averages and variances, respectively and $\sigma_{xy}$ is their covariance. $c_{1},c_{2}$ are small constants that are used for averting division by zero. BTD-HIRLS is again used for comparison purposes, with its regularization parameter being finely tuned in accordance with SSIM. As it can be seen in Fig.~\ref{fig:ssim}, the BBTD algorithm outperforms BTD-HIRLS, exhibiting higher or similar SSIM values over a wide range of spectral bands. 
The CPD model, resulting from the BCPD method, does not capture the low-rank structure of the HSI tensor equally well. %This should be expected in view of the fact that CPD is not a natural model for this type of data. This is also corroborated by the quite large value that BCPD gives as an estimate of the number of endmembers.
The superior performance of the BTD-based algorithms as compared to the CPD one can be explained by the estimated ranks in each case. Specifically, the estimated $\hat{R}$ by both BBTD and BTD-HIRLS is~9. Based on the compelling interpretation that the BTD model offers when it comes to HSI decomposition, $\hat{R}$ corresponds to the number of endmembers (distinct materials) that exist in the depicted scene. That said, $\hat{R}=9$ turns out to be in a good agreement with what is known in the HSI literature for the number of endmembers existing in the scene depicted by the Washington DC Mall AVIRIS HSI~\cite{grk19a}. On the other hand, the BCPD estimate of the CPD rank is $\hat{R}=32$, that is, it largely overestimates the number of endmembers in the scene. It should not be surprising that the CPD model is not able to provide an accurate tensor representation of the Washington DC Mall AVIRIS HSI, manifesting the limitations of the CPD representation in capturing the inherent structure of HSI.

\begin{figure}
\begin{tabular}{c c}
   \includegraphics[width=0.22\textwidth]{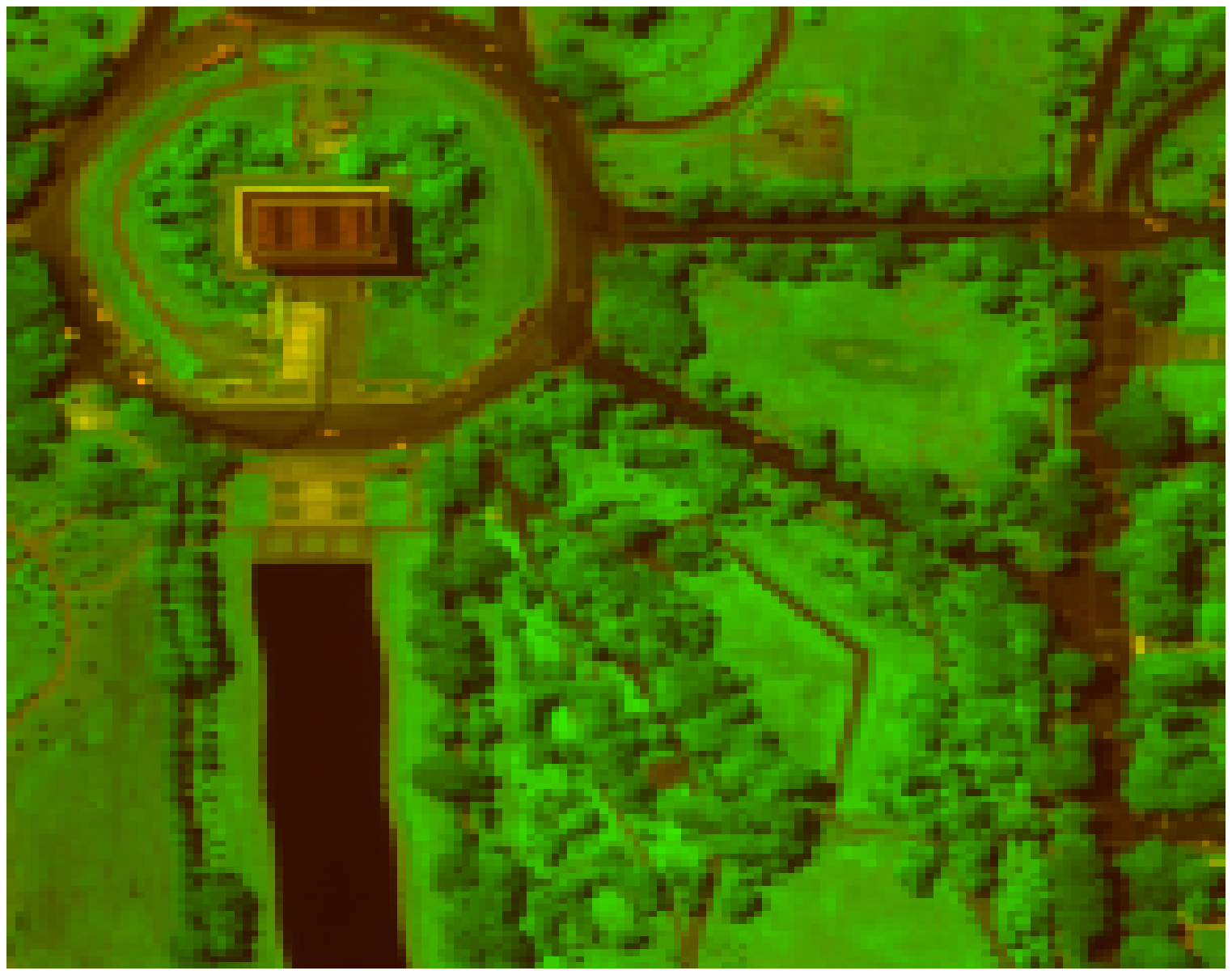} &    \includegraphics[width=0.22\textwidth]{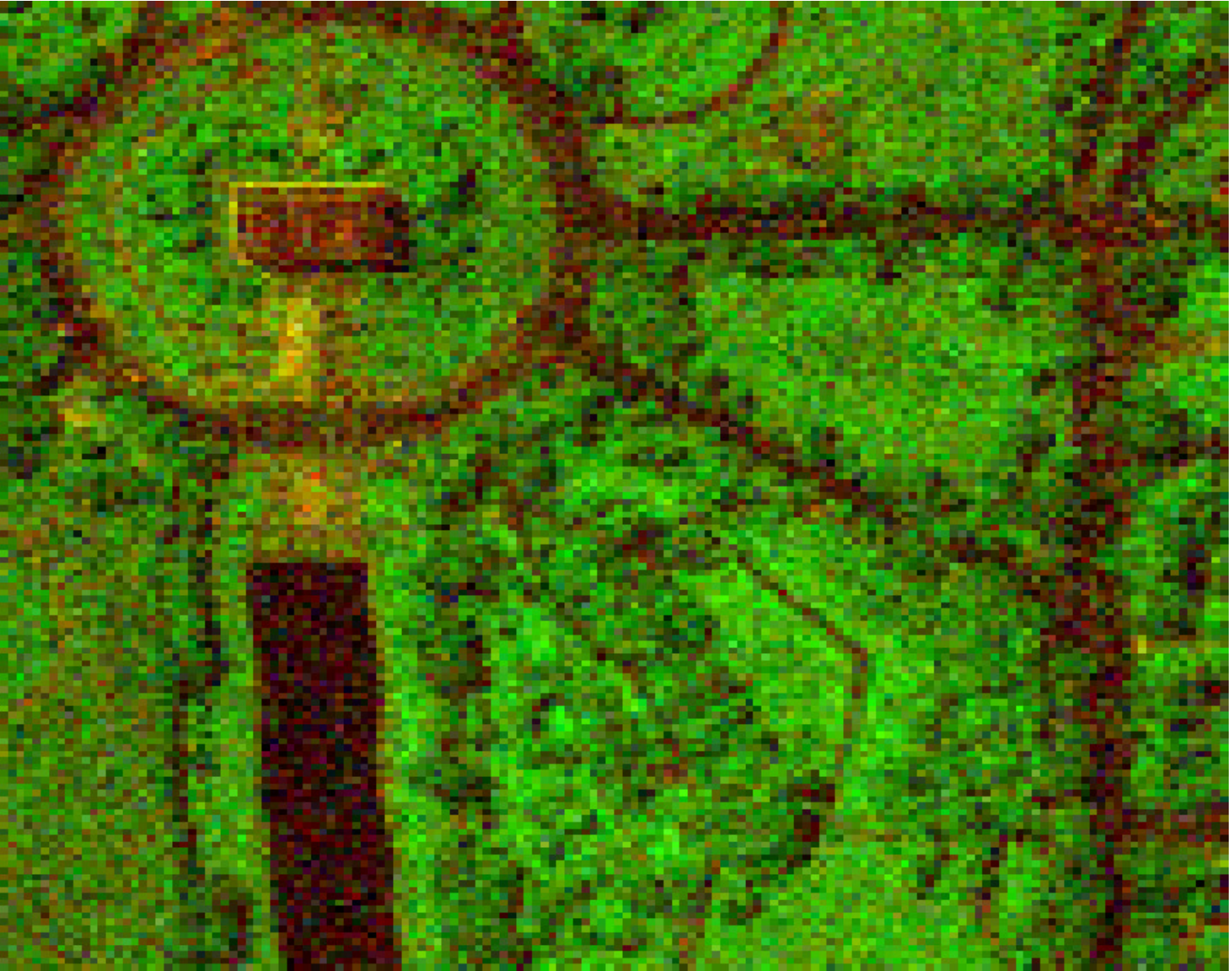} \\
   (a) & (b) \\
      \includegraphics[width=0.22\textwidth]{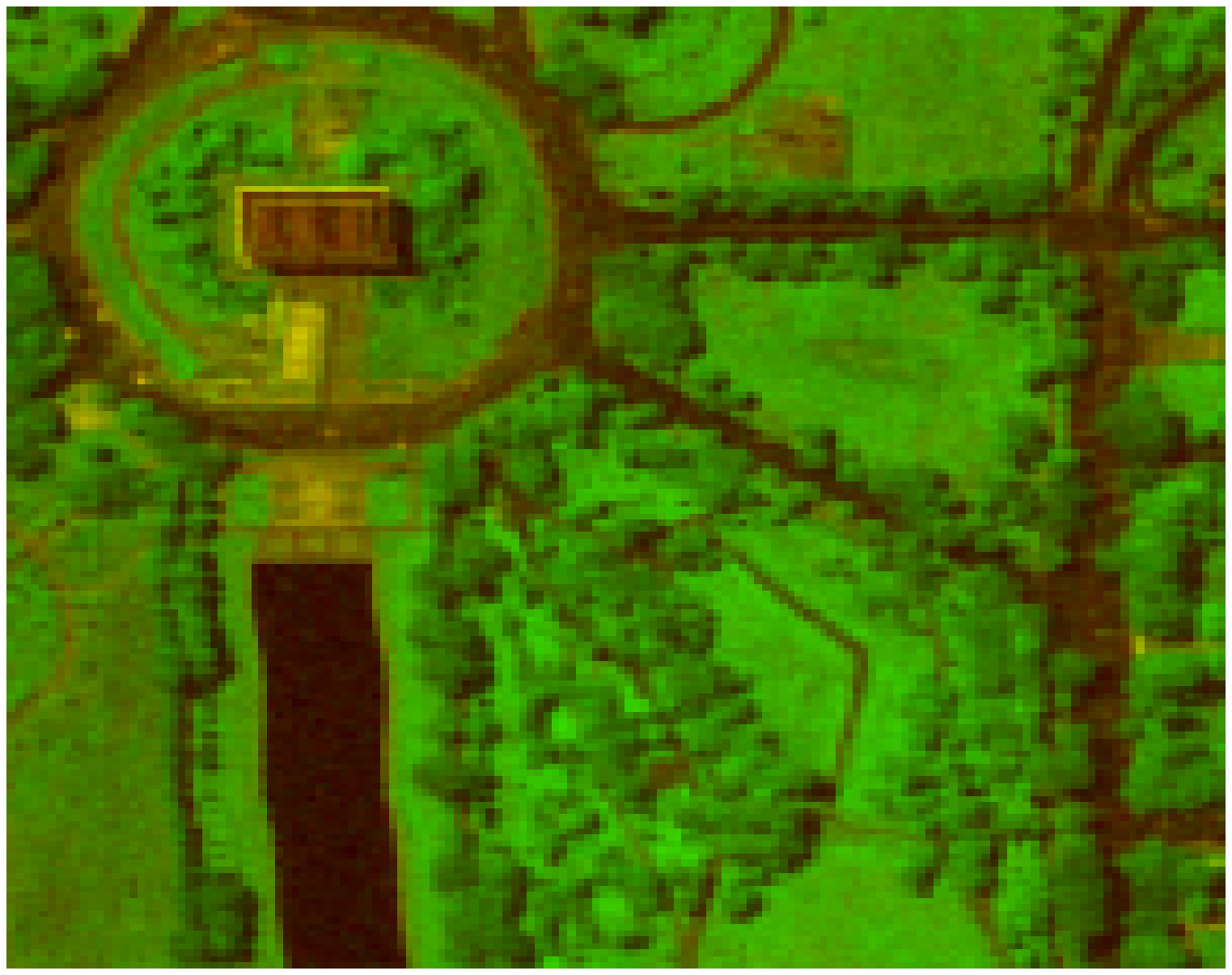} &    \includegraphics[width=0.22\textwidth]{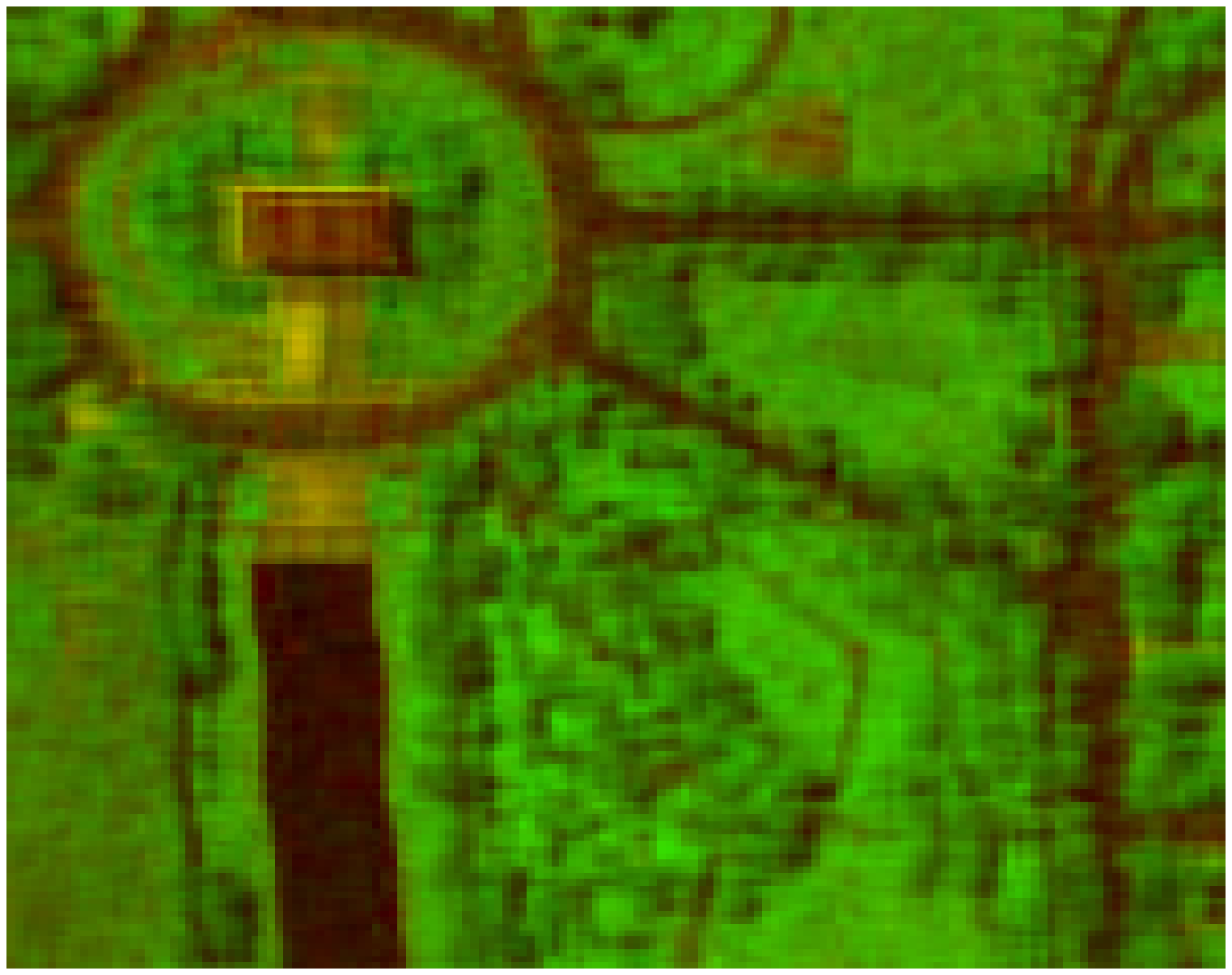} \\
      (c) & (d)
      \end{tabular}
\begin{tabular}{c}
 \hspace{2cm}   \includegraphics[width=0.22\textwidth]{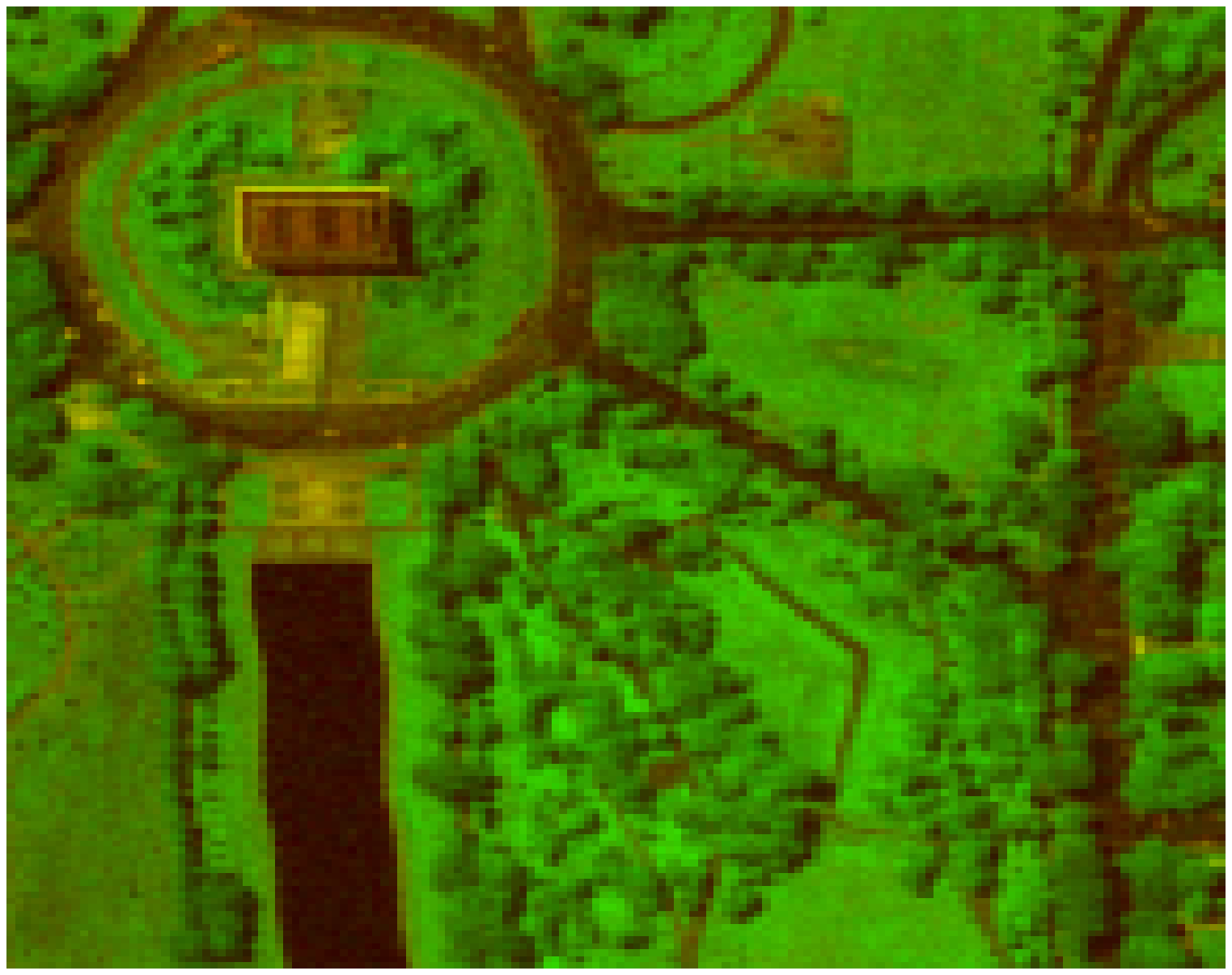} \\
   \hspace{2cm}       (e)
     \end{tabular}
         \caption{False color RGB images (made from bands~34, 64, and~135) of the Washington DC Mall AVIRIS hyperspectral image. (a) Original; (b) Noisy; Denoised with (c) BTD-HIRLS, (d) BCPD, and (e) the proposed BBTD algorithm.}
      \label{fig:hsi_rec}
\end{figure}

For a visual comparison of the results of the three algorithms, Figs.~\ref{fig:hsi_rec}(a) and~(b) depict false color images of the true and the noisy image, respectively, while the BTD-HIRLS, BCPD and BBTD reconstruction results are respectively given in parts~(c), (d) and~(e) of the figure.
The comparable performance of the two BTD methods observed in Fig.~\ref{fig:ssim} is confirmed here by visual inspection. Moreover, as expected, BCPD provides clearly poorer results, with a blurring effect being clearly visible in the corresponding false color image. 

\section{Conclusions}
\label{sec:concls}

As a follow-up to earlier work of ours on BTD model selection and computation based on $\ell_{1,2}$ norm-based regularization, we developed \emph{for the first time} in this paper a Bayesian method for the same problem, which completely relieves its user from having to tune a regularization parameter. The proposed fully-automatic variational inference scheme originates from a Bayesian probabilistic model designed to match perfectly with the BTD model structure and promote model selection through heavy-tailed prior distributions assigned to the BTD factors in the spirit of ARD and SBL. Two alternative simplified Bayesian models were also presented and their model selection properties were investigated by way of their individual joint posterior distribution maximization tasks. Extensive empirical results showed that the proposed algorithm is extremely robust to initialization, converges fast and its model selection ability is comparable to that of its regularization-based counterpart, which however requires parameter fine–tuning. Finally, the appropriateness of the BTD model in approximating hyperspectral imaging data was demonstrated in a HSI denoising experiment, where the proposed algorithm was favorably compared to a Bayesian rank-revealing CPD algorithm emanating from one of the simplified models mentioned above. 

Future work will focus on the development of constrained (e.g., to ensure nonnegativity) and online variants of the proposed method.

\bibliographystyle{IEEEtran}
\bibliography{IEEEabrv,refs}

\begin{thebibliography}{10}
\providecommand{\url}[1]{#1}
\csname url@samestyle\endcsname
\providecommand{\newblock}{\relax}
\providecommand{\bibinfo}[2]{#2}
\providecommand{\BIBentrySTDinterwordspacing}{\spaceskip=0pt\relax}
\providecommand{\BIBentryALTinterwordstretchfactor}{4}
\providecommand{\BIBentryALTinterwordspacing}{\spaceskip=\fontdimen2\font plus
\BIBentryALTinterwordstretchfactor\fontdimen3\font minus
  \fontdimen4\font\relax}
\providecommand{\BIBforeignlanguage}[2]{{%
\expandafter\ifx\csname l@#1\endcsname\relax
\typeout{** WARNING: IEEEtran.bst: No hyphenation pattern has been}%
\typeout{** loaded for the language `#1'. Using the pattern for}%
\typeout{** the default language instead.}%
\else
\language=\csname l@#1\endcsname
\fi
#2}}
\providecommand{\BIBdecl}{\relax}
\BIBdecl

\bibitem{ldl08b}
L.~{De~Lathauwer}, ``Decompositions of a higher-order tensor in block terms ---
  {P}art~{II}: Definitions and uniqueness,'' \emph{{SIAM} J. Matrix Anal.
  Appl.}, vol.~30, no.~3, pp. 1033--1066, 2008.

\bibitem{sdfhpf17}
N.~D. Sidiropoulos \emph{et~al.}, ``Tensor decomposition for signal processing
  and machine learning,'' \emph{{IEEE} Trans. Signal Process.}, vol.~65,
  no.~13, pp. 3551--3582, Jul. 2017.

\bibitem{rkg21}
A.~A. Rontogiannis, E.~Kofidis, and P.~V. Giampouras, ``Block-term tensor
  decomposition: Model selection and computation,'' \emph{{IEEE} J. Sel. Topics
  Signal Process.}, vol.~15, no.~3, pp. 464--475, Apr. 2021.

\bibitem{ldl12}
L.~{De~Lathauwer}, ``Block component analysis: a new concept for blind source
  separation,'' in \emph{Proc.~{LVA/ICA-2012}}, Tel Aviv, Israel, Mar. 2012.

\bibitem{qxzzt17}
Y.~Qian, F.~Xiong, S.~Zeng, J.~Zhou, and Y.~Y. Tang, ``Matrix-vector
  nonnegative tensor factorization for blind unmixing of hyperspectral
  imagery,'' \emph{{IEEE} Trans. Geosci. Remote Sens.}, vol.~55, no.~3, pp.
  1776--1792, Mar. 2017.

\bibitem{hl13}
C.~J. Hillar and L.-H. Lim, ``Most tensor problems are {NP}-hard,'' \emph{J.
  ACM}, vol.~60, no.~6, Nov. 2013, article~45.

\bibitem{tp2001}
M.~E. Tipping, ``{Sparse Bayesian learning and the relevance vector machine},''
  \emph{J. Mach. Learn. Res.}, vol.~1, pp. 211--244, Jun. 2001.

\bibitem{pg08}
T.~Park and G.~Casella, ``{The Bayesian lasso},'' \emph{J. Amer. Stat. Assoc.},
  vol. 103, no. 482, pp. 681--686, Jun. 2008.

\bibitem{n96}
R.~M. Neal, \emph{Bayesian Learning for Neural Networks}.\hskip 1em plus 0.5em
  minus 0.4em\relax Springer, 1996.

\bibitem{mh09}
M.~M{\o}rup and L.~K. Hansen, ``Automatic relevance determination for multi-way
  models,'' \emph{J. Chemometrics}, vol.~23, pp. 352--363, 2009.

\bibitem{zzc15b}
Q.~Zhao, L.~Zhang, and A.~Cichocki, ``{Bayesian CP factorization of incomplete
  tensors with automatic rank determination},'' \emph{{IEEE} Trans. Pattern
  Anal. Mach. Intell.}, vol.~37, no.~9, pp. 1751--1763, Sep. 2015.

\bibitem{tlg08}
D.~G. Tzikas, A.~C. Likas, and N.~P. Galatsanos, ``{The variational
  approximation for Bayesian inference},'' \emph{{IEEE} Signal Process. Mag.},
  pp. 131--146, Nov. 2008.

\bibitem{bca17}
D.~M. Blei, A.~Kucukelbir, and J.~D. McAuliffe, ``Variational inference: {A}
  review for statisticians,'' \emph{J. Amer. Stat. Assoc.}, vol. 112, no. 518,
  pp. 859--877, 2017.

\bibitem{zzzca16}
Q.~Zhao, G.~Zhou, L.~Zhang, A.~Cichocki, and S.-I. Amari, ``Bayesian robust
  tensor factorization for incomplete multiway data,'' \emph{{IEEE} Trans.
  Neural Netw. Learn. Syst.}, vol.~27, no.~4, pp. 736--748, Apr. 2016.

\bibitem{dzlz18}
Y.~Du, Y.~Zheng, K.-C. Lee, and S.~Zhe, ``Probabilistic streaming tensor
  decomposition,'' in \emph{Proc.~{ICDM-2018}}, Singapore, Nov. 2018.

\bibitem{ccswt20}
L.~Cheng, Z.~Chen, Q.~Shi, Y.-C. Wu, and S.~Theodoridis, ``Towards
  probabilistic tensor canonical polyadic decomposition 2.0: Automatic tensor
  rank learning using generalized hyperbolic prior,'' arXiv:2009.02472v1
  [cs.LG], Sep. 2020.

\bibitem{zzc15a}
Q.~Zhao, L.~Zhang, and A.~Cichocki, ``{Bayesian sparse Tucker models for
  dimension reduction and tensor completion},'' arXiv:1505.02343v1 [cs.LG], May
  2015.

\bibitem{s20}
D.~Spencer, ``{Inference and uncertainty quantification for high-dimensional
  tensor regression with tensor decompositions and Bayesian methods},'' Ph.D.
  dissertation, Univ. California Santa Cruz, Jun. 2020.

\bibitem{xcww20}
L.~Xu, L.~Cheng, N.~Wong, and Y.-C. Wu, ``Learning tensor train representation
  with automatic rank determination from incomplete noisy data,''
  arXiv:2010.06564v1 [eess.SP], Oct. 2020.

\bibitem{scp21}
F.~Sedighin, A.~Cichocki, and A.-H. Phan, ``Adaptive rank selection for tensor
  ring decomposition,'' \emph{{IEEE} J. Sel. Topics Signal Process.}, vol.~15,
  no.~3, pp. 454--463, Apr. 2021.

\bibitem{zc21}
Y.~Zhou and Y.-M. Cheung, ``{Bayesian low-tubal-rank robust tensor
  factorization with multi-rank determination},'' \emph{{IEEE} Trans. Pattern
  Anal. Mach. Intell.}, vol.~43, no.~1, pp. 62--76, Jan. 2021.

\bibitem{hz19}
C.~Hawkins and Z.~Zhang, ``Bayesian tensorized neural networks with automatic
  rank selection,'' arXiv:1905.10478v1 [cs.LG], May 2019.

\bibitem{zllcz19}
M.~Zhou, Y.~Liu, Z.~Long, L.~Chen, and C.~Zhu, ``Tensor rank learning in {CP}
  decomposition via convolutional neural network,'' \emph{Signal Process.:
  Image Commun.}, vol.~73, pp. 12--21, Apr. 2019.

\bibitem{bnd14}
S.~D. Babacan, S.~Nakajima, and M.~N. Do, ``Bayesian group-sparse modeling and
  variational inference,'' \emph{{IEEE} Trans. Signal Process.}, vol.~62,
  no.~11, pp. 2906--2921, Jun. 2014.

\bibitem{t20}
S.~Theodoridis, \emph{Machine Learning --- A Bayesian and Optimization
  Perspective}, 2nd~ed.\hskip 1em plus 0.5em minus 0.4em\relax Academic Press,
  2020.

\bibitem{zwlj12}
Z.~Zhang, S.~Wang, D.~Liu, and M.~I. Jordan, ``{EP-GIG priors and applications
  in Bayesian sparse learning},'' \emph{J. Mach. Learn. Res.}, vol.~13, pp.
  2031--2061, Jun. 2012.

\bibitem{grtk17}
P.~V. Giampouras, A.~A. Rontogiannis, K.~E. Themelis, and K.~D. Koutroumbas,
  ``{Online sparse and low-rank subspace learning from incomplete data: A
  Bayesian view},'' \emph{Signal Process.}, vol. 137, pp. 199--212, 2017.

\bibitem{bppc13}
A.~J. Brockmeier, J.~C. Principe, A.-H. Phan, and A.~Cichocki, ``Greedy
  algorithm for model selection of tensor decompositions,'' in
  \emph{Proc.~{ICASSP-2013}}, Vancouver, Canada, May 2013.

\bibitem{grk21a}
P.~V. Giampouras, A.~A. Rontogiannis, and E.~Kofidis, ``A {B}ayesian approach
  to block-term tensor decomposition model selection and computation,''
  arXiv:2101.02931v1 [stat.ME], Jan. 2021.

\bibitem{grk21b}
------, ``A {Bayesian} approach to block-term tensor decomposition model
  selection and computation,'' in \emph{Proc.~{EUSIPCO-2021}}, Dublin, Ireland,
  Aug. 2021.

\bibitem{lc11}
A.~J. Lemonte and G.~M. Cordeiro, ``The exponentiated generalized inverse
  {G}aussian distribution,'' \emph{Stat. Prob. Lett.}, vol.~81, pp.
  506–--517, 2011.

\bibitem{gofzc20}
J.~H. {de~M.~Goulart}, P.~M.~R. {de~Oliveira}, R.~C. Farias, V.~Zarzoso, and
  P.~Comon, ``Alternating group lasso for block-term tensor decomposition and
  application to {ECG} source separation,'' \emph{{IEEE} Trans. Signal
  Process.}, vol.~68, pp. 2682--2696, Apr. 2020.

\bibitem{gtrk16}
P.~V. Giampouras, K.~E. Themelis, A.~A. Rontogiannis, and K.~D. Koutroumbas,
  ``Simultaneously sparse and low-rank abundance matrix estimation for
  hyperspectral image unmixing,'' \emph{{IEEE} Trans. Geosci. Remote Sens.},
  vol.~54, no.~8, pp. 4775--4789, 2016.

\bibitem{xzq19}
F.~Xiong, J.~Zhou, and Y.~Qian, ``Hyperspectral restoration via ${L}_0$
  gradient regularized low-rank tensor factorization,'' \emph{{IEEE} Trans.
  Geosci. Remote Sens.}, vol.~57, no.~12, pp. 10\,410--10\,425, Dec. 2019.

\bibitem{grk19a}
P.~V. Giampouras, A.~A. Rontogiannis, and K.~D. Koutroumbas, ``Alternating
  iteratively reweighted least squares minimization for low-rank matrix
  factorization,'' \emph{{IEEE} Trans. Signal Process.}, vol.~67, no.~2, pp.
  490--503, Jan. 2019.

\end{thebibliography}

\end{document}